\documentclass[a4paper,11pt]{article}

\usepackage{jinstpub} 
\usepackage{color} 
\usepackage{xcolor} 
\usepackage{graphicx} 
\usepackage{stmaryrd}
\usepackage{tikz} 
\usetikzlibrary{shapes,arrows,positioning,automata,backgrounds,calc,er,patterns,decorations.markings, shapes.misc} 
\usepackage[compat=1.1.0]{tikz-feynman} 
\usepackage{circuitikz} 
\ctikzset{bipoles/length=.8cm}
\usepackage{subcaption} 
\usepackage{physics} 
\usepackage{units} 
\usepackage{mathtools} 
\usepackage{cleveref}
\usepackage{wrapfig} 
\usepackage{tabularx}
\usepackage{booktabs}
\usepackage{makecell}
\usepackage{multirow}
\usepackage{multicol}
\usepackage[style=chem-angew,
            maxbibnames=1,
            url=true,
            doi=true,
            isbn=false,
            ]{biblatex}
\AtEveryBibitem{%
  \clearfield{note}%
  \clearlist{language}%
}
\DefineBibliographyStrings{english}{%
  mathesis = {Master's thesis},
}

\newcommand\numberthis{\addtocounter{equation}{1}\tag{\theequation}}

\title{Prototype acoustic positioning system for the Pacific Ocean Neutrino Experiment}

\collaboration{The P-ONE Collaboration}
\author[a]{M.~Agostini,}
\author[b]{S.~Agreda,}
\author[a]{A.~Alexander Wight,}
\author[c,d]{P.~S.~Barbeau,}
\author[b]{A.~J.~Baron,}
\author[e]{S.~Bash,}
\author[e]{C.~Bellenghi,}
\author[b]{B.~Biffard,}
\author[e]{M.~Boehmer,}
\author[e]{M.~Brandenburg,}
\author[b]{P.~Bunton,}
\author[c,d]{N.~Cedarblade-Jones,}
\author[b]{M.~Charlton,}
\author[a]{B.~Crudele,}
\author[f]{M.~Danninger,}
\author[g]{T.~DeYoung,}
\author[h]{F.~Fuchs,}
\author[f]{A.~G\"{a}rtner,}
\author[g]{J.~Garriz,}
\author[f]{D.~Ghuman,}
\author[e]{L.~Ginzkey,}
\author[f]{T.~Glukler,}
\author[e]{V.~Gousy-Leblanc,}
\author[f]{D.~Grant,}
\author[f]{A.~Grimes,}
\author[i]{C.~Haack,}
\author[h]{R.~Halliday,}
\author[b]{M.~Heesemann,}
\author[b]{D.~Hembroff,}
\author[f]{F.~Henningsen,}
\author[b]{J.~Hutchinson,}
\author[j]{S.~Karanth,}
\author[e]{T.~Kerscher,}
\author[j]{K.~Kopa\'{n}ski,}
\author[i]{C.~Kopper,}
\author[f]{P.~Krause,}
\author[k]{C.~B.~Krauss,}
\author[l]{N.~Kurahashi,}
\author[e]{C.~Lagunas Gualda,}
\author[e]{K.~Leism\"{u}ller,}
\author[e]{R.~Li,}
\author[e]{S.~Loipolder,}
\author[h]{A.~Maga\~{n}a Ponce,}
\author[e]{S.~Magel,}
\author[j]{P.~Malecki,}
\author[a]{G.~Marshall,}
\author[k]{T.~Martin,}
\author[f]{C.~Miller,}
\author[k]{N.~Molberg,}
\author[k]{R.~Moore,}
\author[b]{L.~Muzi,}
\author[e]{B.~N\"{u}hrenb\"{o}rger,}
\author[f]{B.~Nichol,}
\author[j]{W.~Noga,}
\author[e]{R.~\O{}rs\o{}e,}
\author[e]{L.~Papp,}
\author[g]{V.~Parrish,}
\author[e]{P.~Pfahler,}
\author[b]{B.~Pirenne,}
\author[b]{E.~Price,}
\author[m,n]{A.~Rahlin,}
\author[k]{M.~Rangen,}
\author[e]{E.~Resconi,}
\author[k]{S.~Robertson,}
\author[g]{D.~Salazar-Gallegos,}
\author[e]{A.~Scholz,}
\author[i]{L.~Schumacher,}
\author[j]{S.~Sharma,}
\author[e]{C.~Spannfellner,}
\author[f]{J.~Stacho,}
\author[o]{I.~Taboada,}
\author[g]{J.~P.~Twagireyezu,}
\author[g]{M.~Un Nisa,}
\author[k]{B.~Veenstra,}
\author[g]{C.~Weaver,}
\author[g]{N.~Whitehorn,}
\author[e]{L.~Winter,}
\author[j]{R.~Wro\'{n}ski,}
\author[k]{J.~P.~Ya\~{n}ez,}
\author[c,d]{A.~Zaalishvili}

\affiliation[a]{University College London, London, United Kingdom}
\affiliation[b]{Ocean Networks Canada, University of Victoria, Victoria, BC, Canada}
\affiliation[c]{Department of Physics, Duke University, Durham, NC, 27708, USA}
\affiliation[d]{Triangle Universities Nuclear Laboratory, Durham, NC, 27708, USA}
\affiliation[e]{Physik-department, Technische Universit\"{a}t M\"{u}nchen, D-85748 Garching, Germany}
\affiliation[f]{Department of Physics, Simon Fraser University, 8888 University Drive Burnaby, B.C. Canada,V5A 1S6}
\affiliation[g]{Department of Physics and Astronomy, Michigan State University, East Lansing, MI 48824, USA}
\affiliation[h]{Department of Physics, Elmhurst University, 190 S. Propsect Ave, Elmhurst, IL, 60126, USA}
\affiliation[i]{Erlangen Centre for Astroparticle Physics, Friedrich-Alexander-Universit{\"a}t Erlangen-N\"{u}rnberg, D-91058 Erlangen, Germany}
\affiliation[j]{Institute of Nuclear Physics, Polish Academy of Sciences, Kraków, Poland}
\affiliation[k]{Department of Physics, University of Alberta, Edmonton, Alberta, Canada, T6G 2E1}
\affiliation[l]{Department of Physics, Drexel University, 3141 Chestnut Street, Philadelphia, PA, 19104, USA}
\affiliation[m]{Department of Astronomy and Astrophysics, University of Chicago, 5640 South Ellis Avenue, Chicago, IL, 60637, USA}
\affiliation[n]{Kavli Institute for Cosmological Physics, University of Chicago, 5640 South Ellis Avenue, Chicago, IL, 60637, USA}
\affiliation[o]{School of Physics and Center for Relativistic Astrophysics, Georgia Institute of Technology, Atlanta, GA 30332, USA}

\emailAdd{felix\_henningsen@sfu.ca}
\emailAdd{dilraj\_ghuman@sfu.ca}

\abstract{
We present the design and initial performance characterization of the prototype acoustic positioning system intended for the Pacific Ocean Neutrino Experiment. It comprises novel piezo-acoustic receivers with dedicated filtering- and amplification electronics installed in P-ONE instruments and is complemented by a commercial system comprised of cabled and autonomous acoustic pingers for sub-sea installation manufactured by Sonardyne Ltd. We performed an in-depth characterization of the acoustic receiver electronics and their acoustic sensitivity when integrated into P-ONE pressure housings. These show absolute sensitivities of up to \unit[-125]{dB re V$^2/$$\mu$Pa$^2$} in a frequency range of \unit[$10-40$]{kHz}. \textcolor{rev1}{We furthermore conducted a positioning measurement campaign in the ocean by deploying three autonomous acoustic pingers on the seafloor, as well as a cabled acoustic interrogator and a P-ONE prototype module deployed from a ship}. Using a simple peak-finding detection algorithm, we observe high accuracy in the tracking of \textcolor{rev1}{relative ranging times at approximately \unit[$230-280$]{$\mu$s} at distances of up to \unit[1600]{m}, which is sufficient for positioning detectors in a cubic-kilometer detector and which can be further improved with more involved detection algorithms}. The tracking accuracy is further confirmed by independent ranging of the Sonardyne system and closely follows the ship's drift in the wind measured by GPS. The absolute positioning shows the same tracking accuracy with its absolute precision only limited by the large uncertainties of the deployed pinger positions on the seafloor.
}


\usepackage{lineno}

\definecolor{rev1}{rgb}{0,0,0}
\definecolor{rev2}{rgb}{0,1,0}
\definecolor{rev3}{rgb}{0,0,1}

\begin{document}
\maketitle
\flushbottom

\section{Introduction}
\label{sec:intro}
The Pacific Ocean Neutrino Experiment (P-ONE)~\cite{agostini_pacific_2020} is a future Neutrino Telescope that will instrument a volume of more than a cubic kilometer in the deep ocean with photosensor instruments. Located on the Cascada Basin abyssal plain, off the coast of Vancouver Island in Canada, this three-dimensional array of sensors will allow the detection of high-energy neutrinos~\cite{henningsen_pacific_2023}.

Photomultiplier tubes (PMT) and electronics for P-ONE optical modules (P-OM)~\cite{spannfellner_design_2023} and calibration modules (P-CALs)~\cite{stacho_development_2023} are encapsulated in 17-inch glass spheres and attached to 1-kilometer-long mooring lines. These moorings act as the mechanical interface for the instruments, carry all electrical and fiber connections from the instruments to the seafloor infrastructure, and are kept upright with subsea floats. For precision directional reconstruction of neutrinos, continuous knowledge of the detector position is critical and requires a continuous tracking system for the detector geometry. 

The moving lines in the ocean are pulled by the dynamic environment of the ocean currents and will move with time. This variation of the detector geometry requires a continuous positioning system for the P-ONE instruments and, subsequently, its mooring lines. \textcolor{rev1}{Given the speed of light in water, the approximate \unit[1]{ns} resolution of P-ONE PMTs requires a relative positioning resolution of \unit[20]{cm} or better. Distributed systems and specialized detection algorithms offer ways to reduce the positioning error even below this limit in order to make it a sub-dominant contribution to photon timing uncertainties}.

\begin{figure}[h!]
    \centering
    \begin{subfigure}{.44\textwidth}
      \centering
      \includegraphics[width=0.9\textwidth]{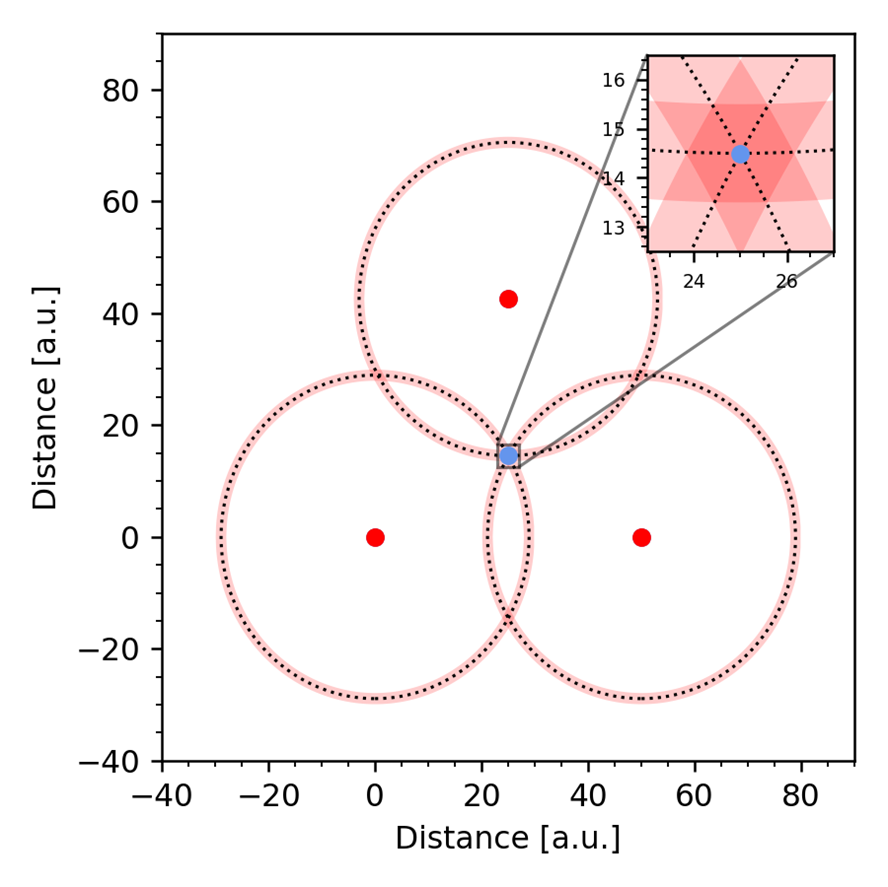}
      \caption{Two-dimensional multilateration.}
      \label{fig:system-concept-a}
    \end{subfigure}%
    \begin{subfigure}{.44\textwidth}
      \centering
      \includegraphics[width=0.9\textwidth]{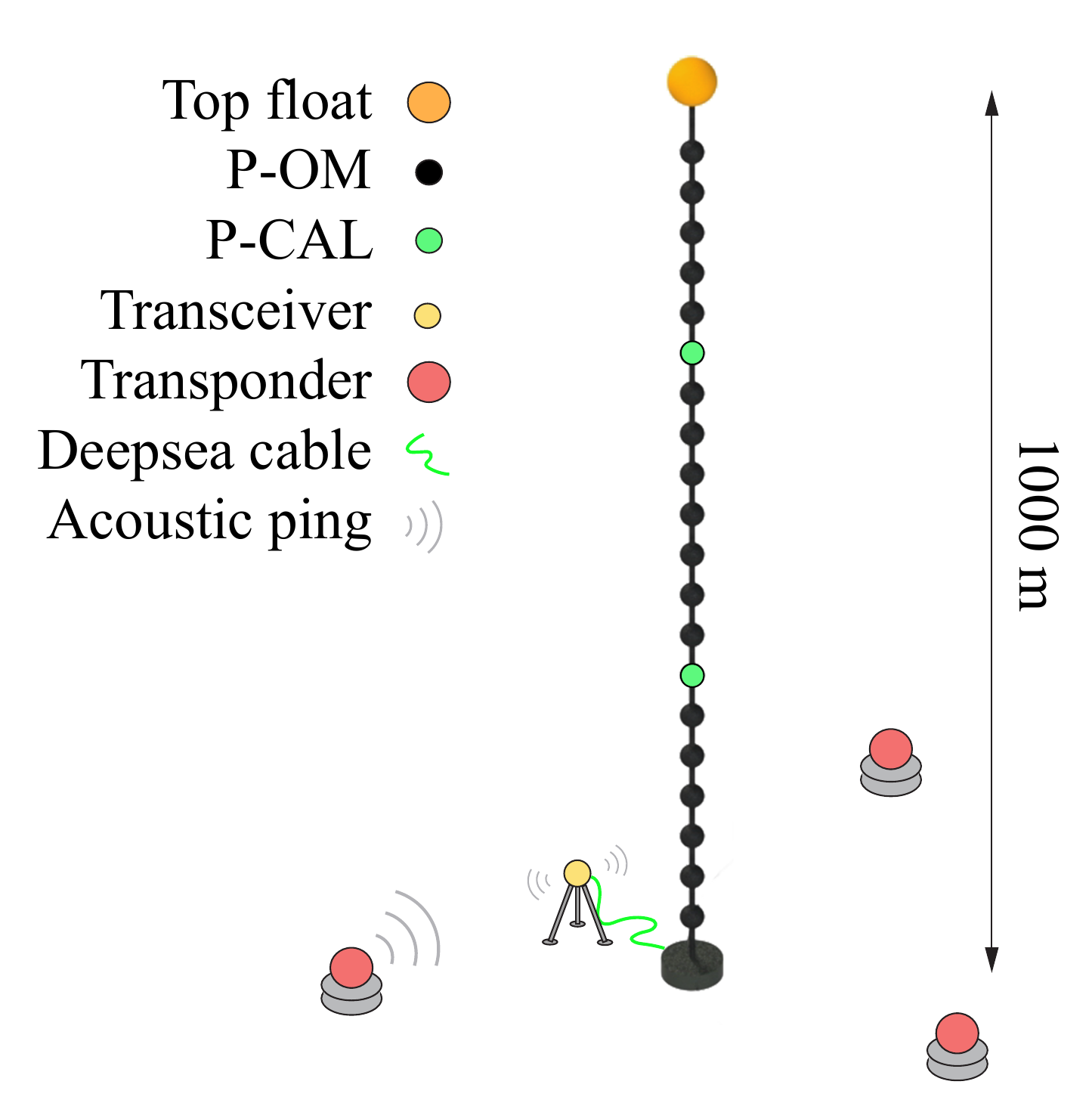}
      \caption{P-ONE acoustic calibration seafloor system.}
      \label{fig:system-concept-b}
    \end{subfigure}
    \caption{\textbf{a)} Multilateration in two dimensions using three known locations (red) and a receiving location or point of interest (blue). Arrival times are used to measure the distance between all positions and ultimately allow identifying the location of the point of interest. Shaded areas illustrate uncertainties in the distance measurement. Figure taken from~\cite{ghuman_acoustic_2023}. \textbf{b)} P-ONE acoustic positioning system using three acoustic transponders or beacons (red), an acoustic transceiver (yellow) and piezo-acoustic receivers in P-OMs (black) and P-CALs (green). Figure from~\cite{park_canadian_2025}; modified.}
    \label{fig:system-concept}
\end{figure}

Deep-ocean positioning systems are established in the domain of marine science and industry and are especially prominent in deep-sea exploration and submersibles. They most commonly use acoustic signals, with systems composed of multiple emitters and receivers, measuring the time of flights (TOFs) between known acoustic beacon locations within the system and the receiver as the point of interest. These TOFs can be converted to ranges with knowledge of the propagation of acoustic signals in ocean water, and multilateration is used to combine them and estimate the position of interest in three-dimensional space. This reconstructed position is valid within the reference coordinate system spanned by the known beacon locations. A two-dimensional depiction of this procedure is shown in \cref{fig:system-concept-a}.

\par\noindent

Several large-volume experiments use positioning systems based on acoustic signals, including KM3NeT~\cite{viola_acoustic_2016} and the IceCube Upgrade~\cite{heinen_acoustic_2021}. In P-ONE we plan to use a system based on piezo-acoustic receivers in every P-OM and P-CAL hemispheres for acoustic positioning~\cite{ghuman_acoustic_2023}. With multiple acoustic beacons emitting signals on the seafloor, piezo-acoustic receivers will provide excellent reconstruction prospects of the mooring line position as a function of time. The planned system foresees multiple acoustic beacons within the instrumented volume of P-ONE. 

The network of acoustic beacons, shown in \cref{fig:system-concept-b}, provides multiple distance baselines for each piezo-acoustic receiver. Measuring acoustic signal TOFs between each known beacon position and a particular piezo-acoustic receiver allows for finding its position within the beacon coordinate system via multilateration. The relative coordinate system spanned by the beacons ultimately needs to be placed in the GPS reference frame. This is achieved in collaboration with Ocean Networks Canada and the Northern Cascadia Subduction Zone Observatory (NCSZO)~\cite{farrugia_northern_2019,hutchinson_initial_2022}. Using an acoustic wave glider, the beacon position can be found with centimeter precision. The mooring line position accuracy is thus primarily determined by the capabilities to reconstruct the acoustic signals propagating between beacons and piezo-acoustic receivers, and their true TOF. 

The first P-ONE mooring, P-ONE-1, will be deployed in the summer of 2025. Its positioning system comprises three autonomous acoustic beacons on the seafloor and a cabled acoustic transceiver, all from the Sonardyne portfolio~\cite{noauthor_sonardyne_nodate}. The transceiver is able to communicate and control the autonomous beacons acoustically. Triggers are sent to the cabled unit and time-stamped with high-precision, and all autonomous beacons respond acoustically after a known turn-around time. Acoustic pings in this system are expected in the frequency ranges of \unit[19 - 40]{kHz}. 

\section{Piezo-acoustic receiver design}
\label{sec:design}
\begin{figure}[h!]
    \centering
    \begin{subfigure}{.575\textwidth}
      \centering
      \includegraphics[width=.7\textwidth]{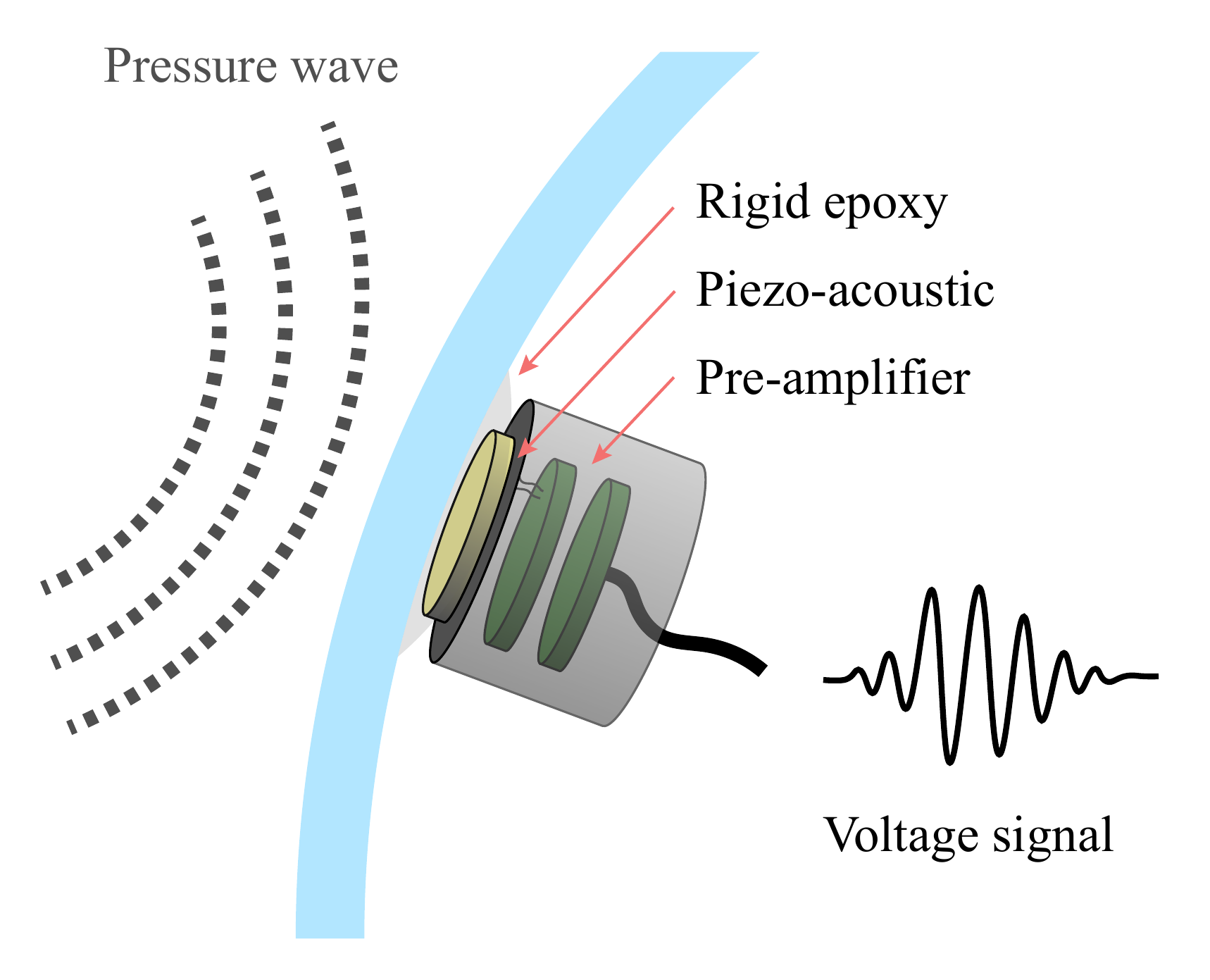}
      \caption{Acoustic receiver unit and detection concept.}
      \label{fig:design-concept-a}
    \end{subfigure}%
    \hspace{2mm}
    \begin{subfigure}{.375\textwidth}
      \centering
      \includegraphics[width=.7\textwidth]{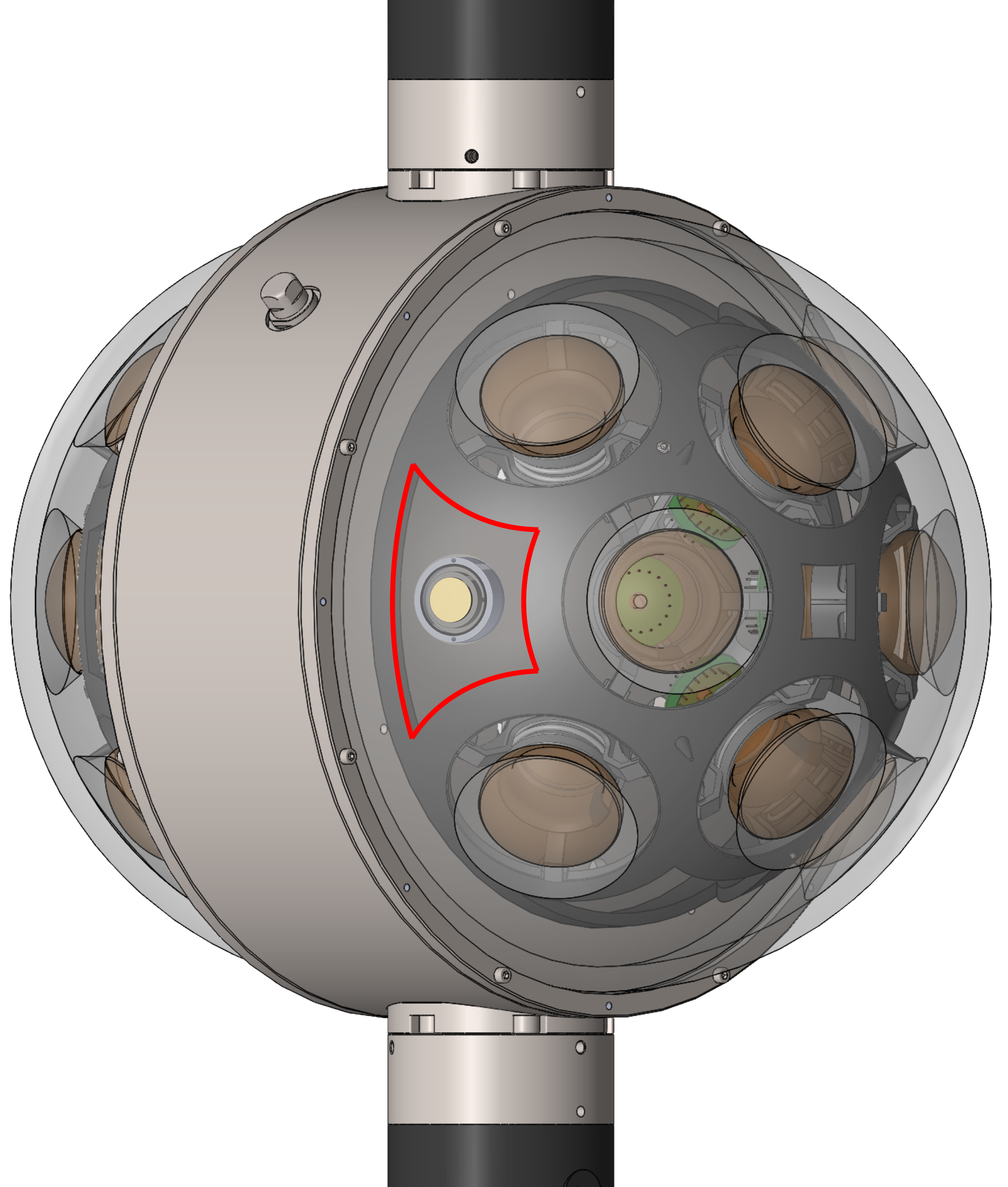}
      \caption{P-ONE optical module.}
      \label{fig:design-concept-b}
    \end{subfigure}
    \caption{\textbf{a)}~Detection concept of external pressure waves with behind-glass acoustic sensors. The instrument pressure housing acts as the resonator and transmitter of external acoustic waves. A piezo-acoustic disk is glued to the glass and responds to the transmitted vibration with voltage signals. Using pre-amplifiers, these small voltage signals can be amplified,  digitized and used for further processing. Figure reproduced from~\cite{park_canadian_2025}. \textbf{b)}~P-ONE optical module (P-OM) with an integrated acoustic receiver. The frame mounting interface is highlighted in red. The main cable proceeds up and down from the Titanium ring (black).}
    \label{fig:design-concept}
\end{figure}

Commercial hydrophones for deep-sea acoustic signal detection use encapsulation materials matching the acoustic impedance of the ocean water to optimize their sensitivity. For P-ONE instruments, acoustic receivers must be installed inside the instrument's glass pressure housing to keep to the connector-less design of P-ONE. Acoustic signals must thus be picked up via the glass pressure housing, so acoustic coupling to the glass is critical. The whole pressure housing of the instrument acts as a resonator and transmits external acoustic pressure waves. Although amplitudes are expected to be dampened with this technique, it is successfully used in KM3NeT~\cite{viola_acoustic_2016} and IceCube~\cite{heinen_acoustic_2021}. An illustration of this concept is shown in \cref{fig:design-concept-a} while the generic P-OM pressure housing design is shown in  \cref{fig:design-concept-b}. Two acoustic receivers are planned to be installed in every P-ONE module. 
%

\subsection{Mechanical layout}
The mechanical design of the receiver unit consists of the piezo-acoustic element, the non-conductive Delrin backing onto which it is mounted together with the analog filtering and amplification electronics, the aluminum housing in which this resides, and the epoxy that attaches this unit to the inside of the glass pressure housing. 
We use the piezo-acoustic element PIC255 from PI Ceramic with a wrapped-electrode configuration This has an outer diameter of \unit[16]{mm} and a thickness of \unit[2]{mm}, and exhibits longitudinal and radial resonance frequencies of \unit[1]{MHz} and \unit[125]{kHz}, respectively~\cite{noauthor_pi_nodate}. These resonance frequencies are far outside our signal frequency band.

\begin{figure}[h!]
    \centering
        \begin{subfigure}{.47\textwidth}
      \centering
      \includegraphics[width=.8\textwidth]{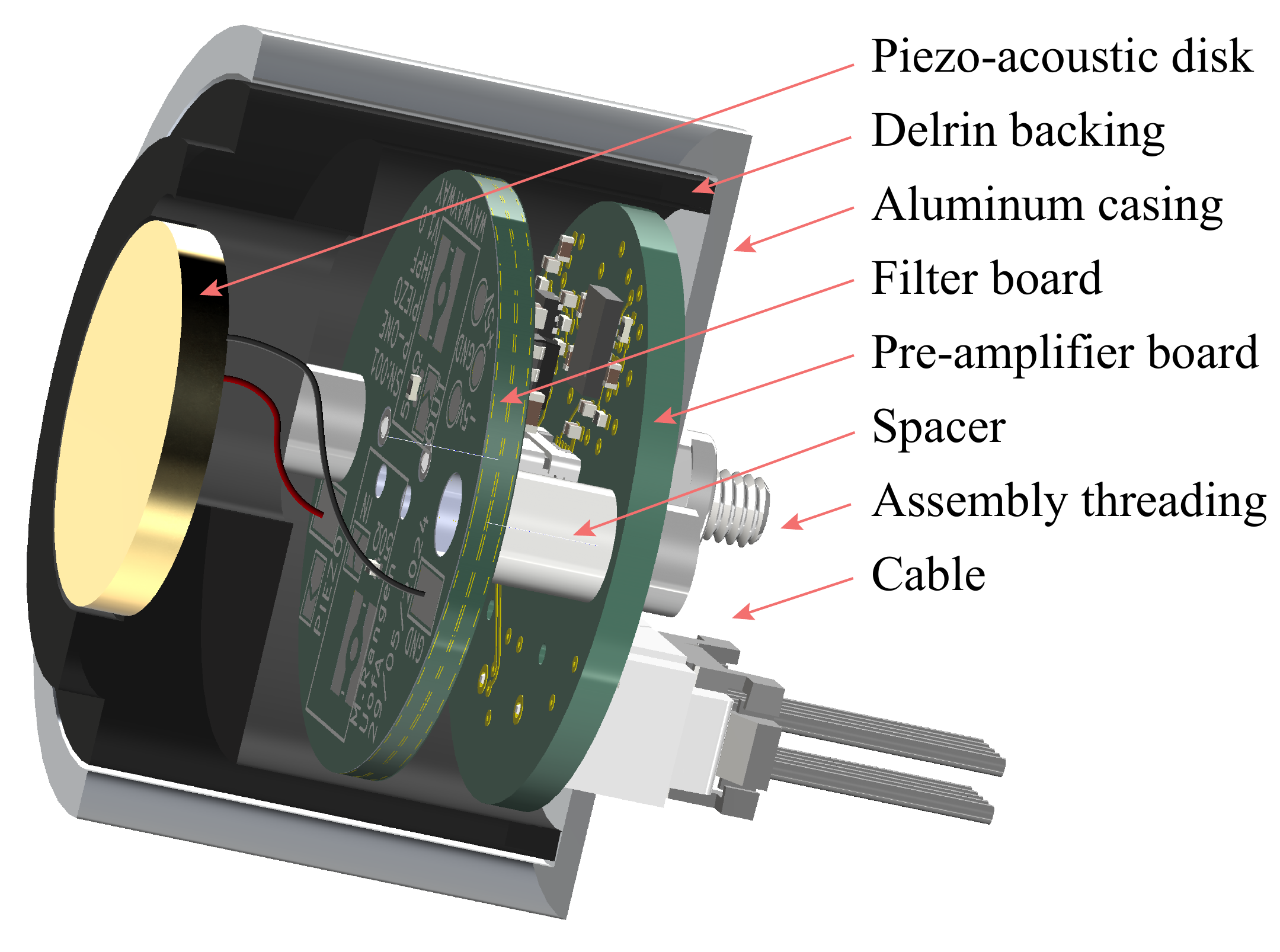}
      \caption{Acoustic receiver unit in section view.}
      \label{fig:section-a}
    \end{subfigure}%
    \hspace{2mm}
    \begin{subfigure}{.47\textwidth}
      \centering
      \includegraphics[width=.8\textwidth]{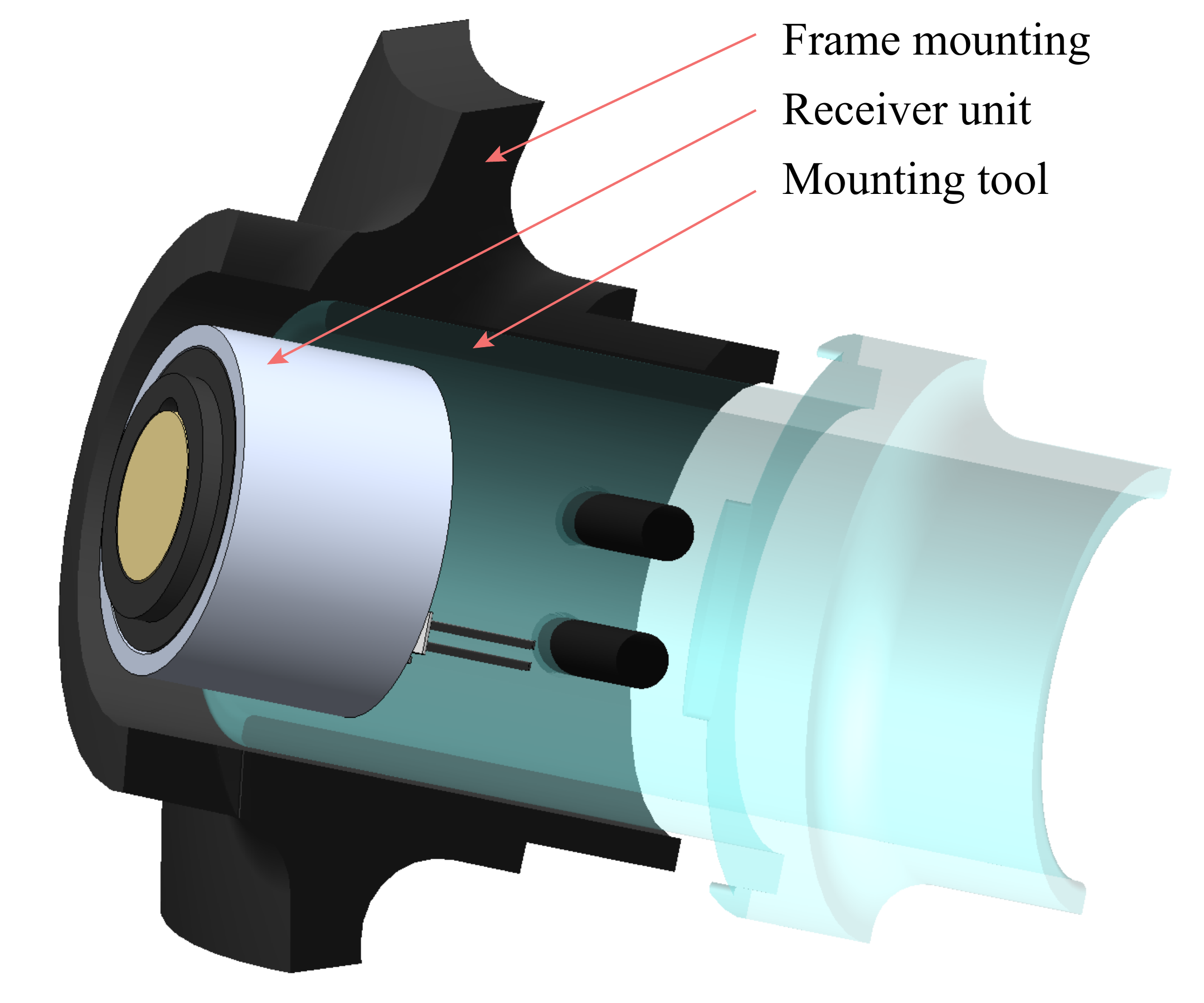}
      \caption{Acoustic receiver with frame insert.}
      \label{fig:section-b}
    \end{subfigure}
    \caption{\textbf{a)}~Mechanical layout of the P-ONE acoustic receiver unit. The image shows the the aluminum housing, the Delrin backing, the piezo-acoustic disk, the filter and pre-amplifier electronic boards, spacers and the cable. \textbf{b)}~P-ONE optical module frame insert for the acoustic receiver unit. The image shows the acoustic receiver unit and the frame mounting insert. The positioning tool is also indicated (blue) but is only used as a guiding piece while gluing and removed afterwards.}
    \label{fig:section}
\end{figure}

To install the receivers we first mount an insert onto the P-ONE module frame. Then, with a drop of rigid epoxy on the piezo disk, we place the receiver unit into its position using a custom tool. The tool and the insert act as installation guides during the curing process. After the epoxy has cured, we remove the positioning tool  and two horizontal screws close the cavity of the frame insert. No additional mechanical coupling is used to keep the acoustic receiver isolated from the system. The frame mount provides protection to internal P-ONE module components in the unexpected scenario of the epoxy detaching from the glass. A schematic view of this assembly is shown in \cref{fig:section} with a detailed view of the receiver and its position within the frame insert. 
\par\noindent
The dimensions of all components in the unit need to be optimized between the curvature of the glass for mechanical coupling and the necessary space requirements for the piezo-acoustic disk and integrated analog electronics. The mechanical design of the Delrin backing and the aluminum shielding is based on the selected piezo element. These pieces were designed to provide a minimal, non-conductive interface to the piezo disk while providing space for printed circuit boards (PCBs) with reasonable size. Here, the epoxy interface to the glass was minimized as much as possible, in order to focus the vibrational signal pickup on the piezo element. Finally, the aluminum housing provides a lightweight shield against electromagnetic interference (EMI), expected from various electronics within the P-OM and which would decrease our readout signal-to-noise ratio (SNR). 
The assembled receiver unit is then glued to the glass using epoxy and the positioning tool as shown in \cref{fig:section-b}.

\subsection{Signal filtering and amplification}
\begin{figure}[h!]
    \centering
    \includegraphics[width=.99\textwidth]{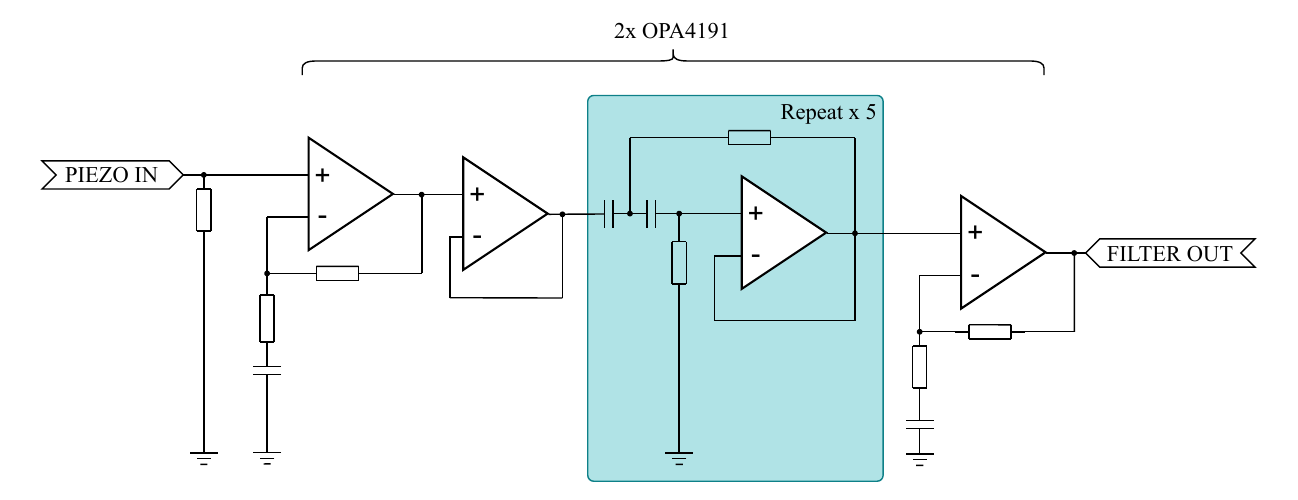}
    \caption{Conceptual schematic of the piezo-acoustic filter board with a first voltage amplification. The second stage decouples the amplification stage impedance before a 5-stage active high-pass filter with a roll-off frequency of \unit[10]{kHz}. Two quad-channel OPA4191 amplifiers are used to realize this.}
    \label{fig:sch-filter}
\end{figure}

The external pressure wave amplitude determines the voltage response of the piezo-acoustic element. In most realistic scenarios, this will result in signals smaller than \unit[1]{mV} and amplification is required for digitization. The electrodes of the piezo-acoustic element are directly coupled to the first of two PCBs in our analog electronics stack. This filter board provides a first voltage amplification and a 5-stage active high-pass filter with a roll-off frequency at \unit[10]{kHz}, which removes low-frequency noise impacting our SNR. This circuit is realized using two quad-channel amplifiers by Texas Instruments (OPA4191), where the second stage is used as a voltage follower to decouple the amplification stage impedance, and the five intermediate stages are used as active high-pass filters. All stages are operated with \unit[$\pm$ 5]{V} dual-supply voltage rails. This concept is shown in \cref{fig:sch-filter}.
\par
\begin{figure}[h!]
    \centering
    \vspace{-12pt}
    \includegraphics[width=.8\textwidth]{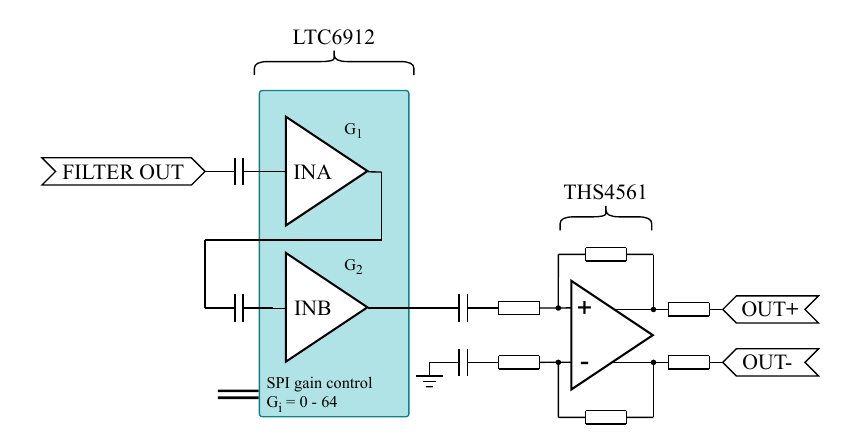}
    \caption{Conceptual schematic of the piezo-acoustic pre-amplifier board. The output of the filter stage is coupled to the gain-adjustable amplifier LTC6912 with two gain stages, and controlled via SPI. A differential conversion of the amplified signal is achieved in a subsequent stage with the THS4561 amplifier. This last step reduces pick-up noise from EMI during signal transmission via unshielded cables to the P-ONE digitization electronics.}
    \label{fig:sch-preamp}
\end{figure}

\noindent
The second PCB in the stack is an inverting pre-amplifier with serial-programmable gain and a differential conversion amplifier for signal output. The programmable-gain amplifier (PGA) is the LTC6912 from Analog Devices, and the differential conversion is realized using the THS4561 by Texas Instruments. The gain of each stage on the amplifier can have eight different values from $0 - 64$, configurable through a serial peripheral interface (SPI). This results in a maximum controllable gain of \unit[$G'_{\text{max}} = 4096$]{V/V}. Including the filter stage and the differential conversion, this means a maximum system gain of \unit[$G\sim 60 \times G'$]{V/V} can be obtained from piezo-acoustic element to signal output. The maximum usable gain of the pre-amplifier stage is ultimately limited by internal and external noise contributions to the system. A differential signal output is used since unshielded single-wire cables are used for transmission to the P-ONE mainboard. Differential transmission is beneficial since pick-up noise from EMI on the unshielded cables is assumed common between both lines and thus reduced when digitized differentially.  All stages are operated with a \unit[5]{V} single-supply voltage rail. A conceptual schematic is shown in \cref{fig:sch-preamp}.
\par\noindent
%

\subsection{Digitization}
The P-ONE data acquisition hosts a dedicated analog-to-digital converter (ADC) for digitizing the acoustic signals of the integrated piezo-acoustic receivers. For this we use the TLV320ADC6140 by Texas Instruments. For P-ONE, this ADC is included with the custom P-ONE mainboard electronics and is integrated into the precision timing of the P-ONE detector network. For acoustic ocean field-testing and performance evaluation, we set up a a prototype acoustic module, described in \cref{subsec:prototype}. 

\begin{figure}[h!]
    \centering
    \includegraphics[width=.94\textwidth]{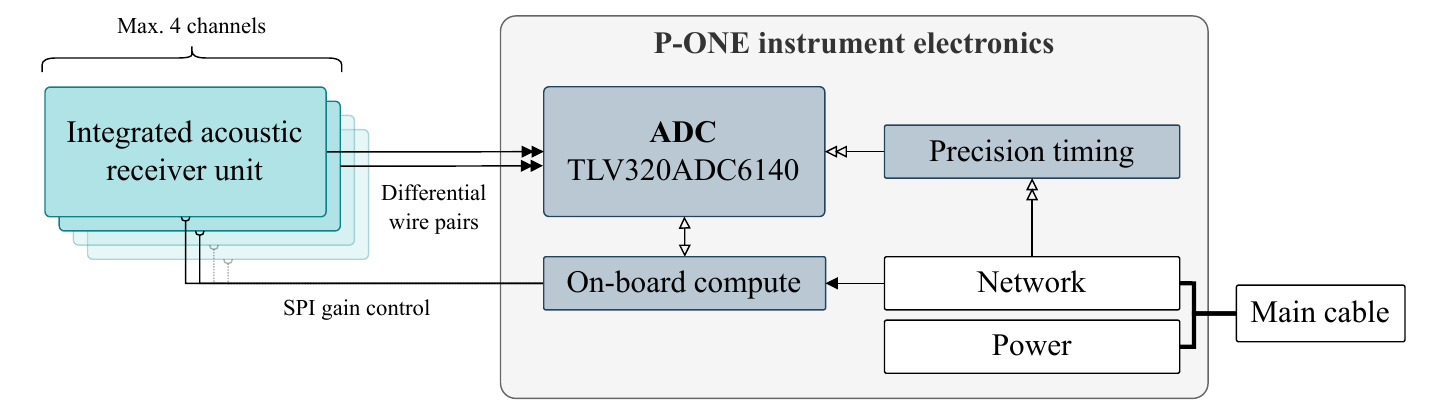}
    \caption{Logic diagram of the data acquisition concept for the P-ONE acoustic receiver readout.}
    \label{fig:adc}
\end{figure}

In P-ONE, the ADC itself will be operated in differential input mode and with sampling rates between \unit[192 - 768]{kHz}. Given the Nyquist-Shannon sampling theorem~\cite{nyquist_certain_1928,shannon_communication_1949} and our signal range of interest between \unit[19-50]{kHz}, even the slowest sampling rate is sufficient for our purposes. The acoustic data taking is expected to run continuously during detector operation, but also custom scheduling of data taking periods, file flagging, and other operational features are supported by the detector data acquisition system. A logic diagram of the P-ONE digitization hardware is given in \cref{fig:adc} and shows all connections of the system. 
%
\subsection{Prototype acoustic module}
\label{subsec:prototype}
\begin{figure}[h!]
    \centering
        \begin{subfigure}{.31\textwidth}
      \centering
      \includegraphics[width=.9\textwidth]{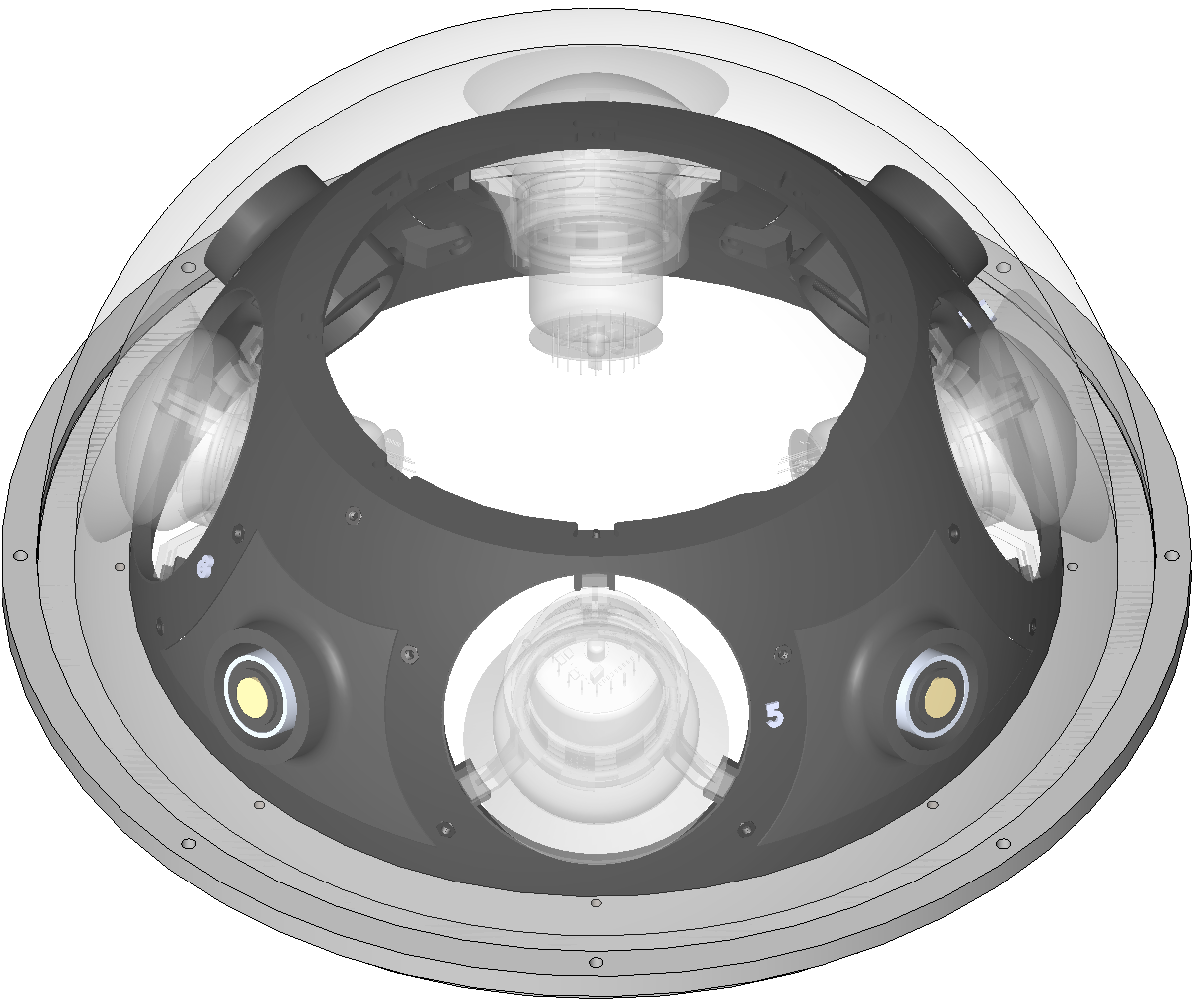}
      \caption{Prototype module design.}
      \label{fig:prototype-a}
    \end{subfigure}%
    \hspace{1.5mm}
    \begin{subfigure}{.39\textwidth}
      \centering
      \includegraphics[width=.9\textwidth]{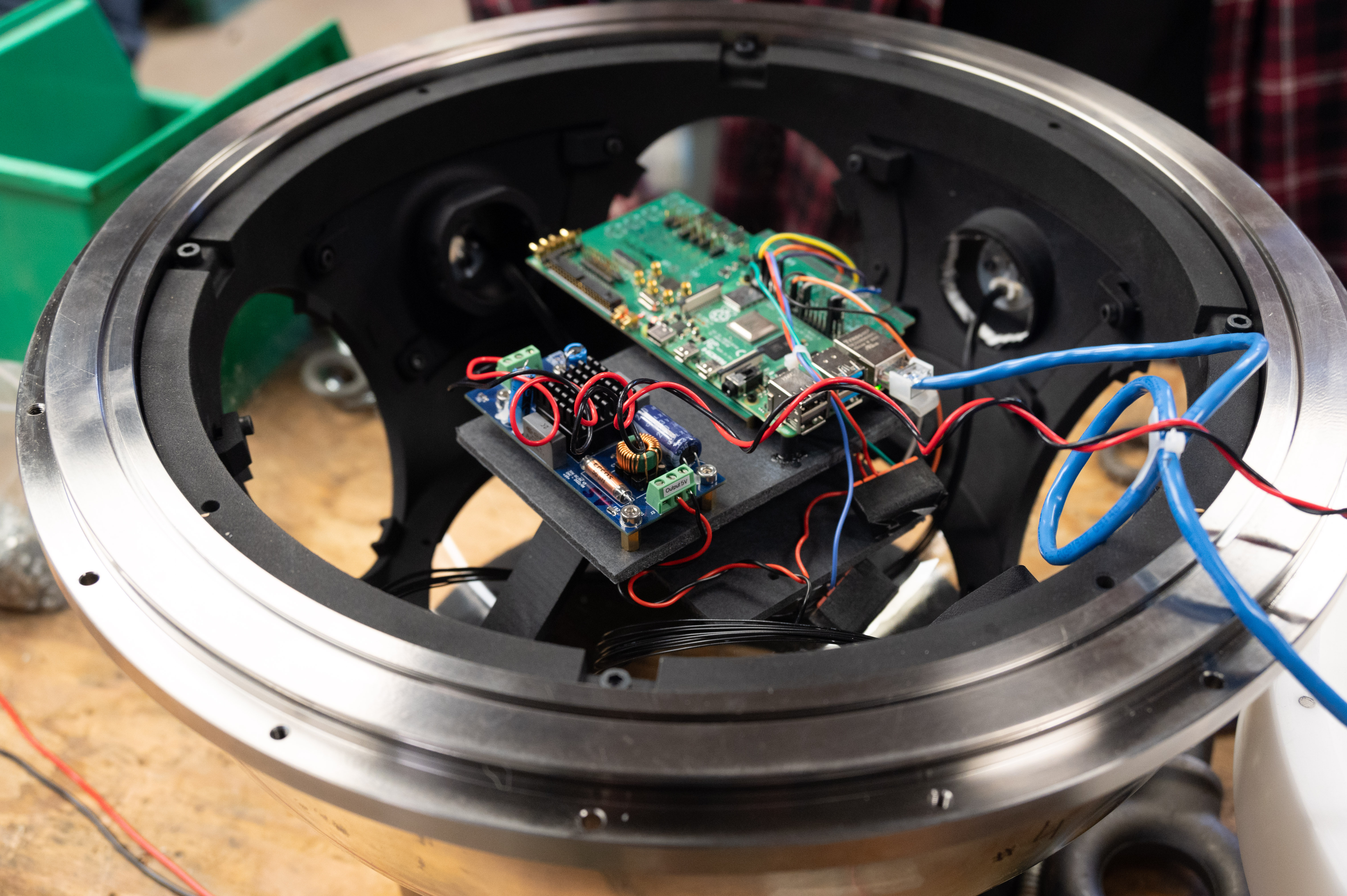}
      \caption{Internal electronics (top view).}
      \label{fig:prototype-b}
    \end{subfigure}
    \hspace{1.5mm}
    \begin{subfigure}{.26\textwidth}
      \centering
      \includegraphics[width=.9\textwidth]{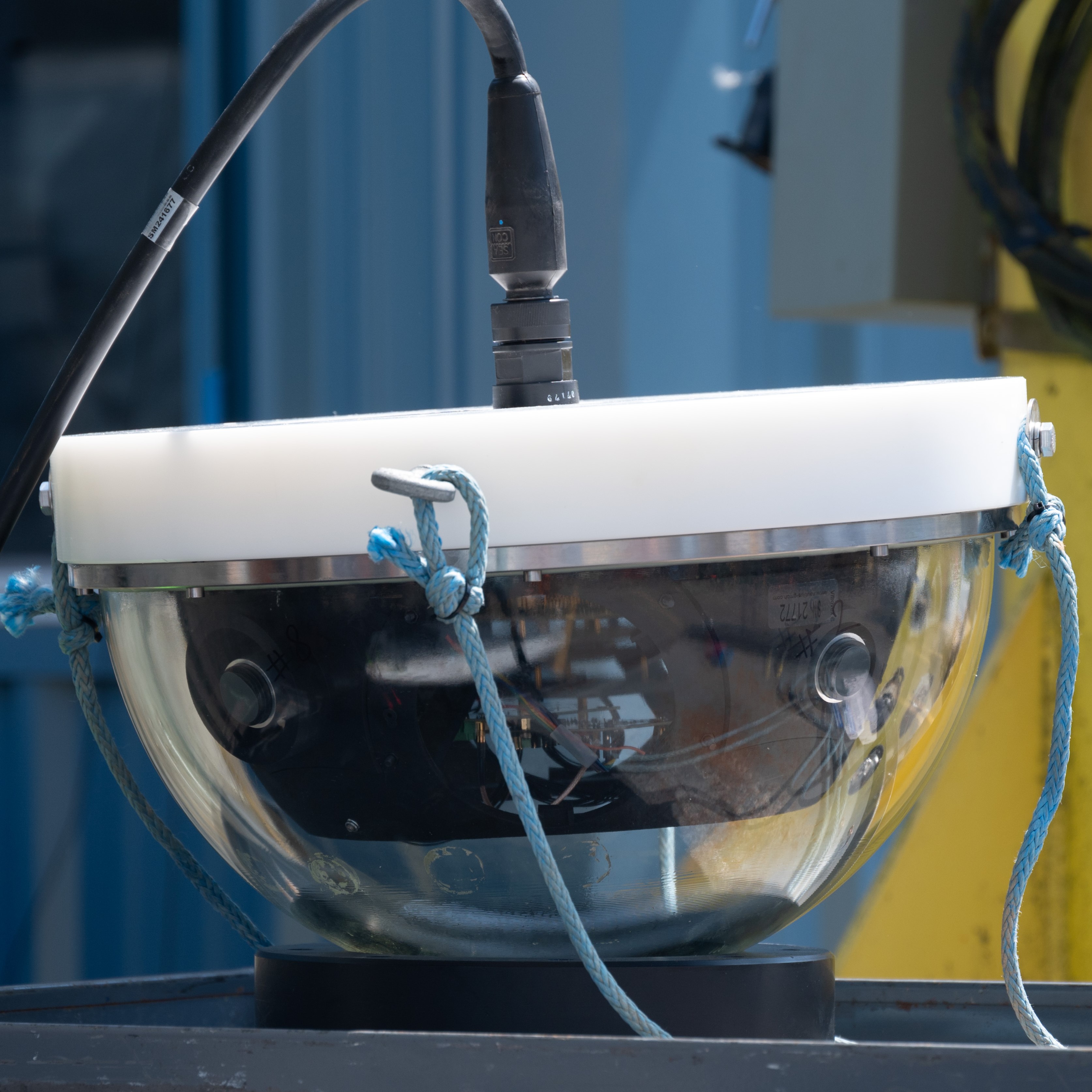}
      \caption{Prototype module.}
      \label{fig:prototype-c}
    \end{subfigure}
    \caption{\textbf{a)}~Computer-animated layout of the prototype acoustic module without internal electronics. \textbf{b)}~Photograph of the internal electronics stack. \textbf{c)}~Photograph of the assembled prototype.}
    \label{fig:prototype}
\end{figure}

For field testing and verification of the P-ONE acoustic receiver design and digitization, we built an acoustic module with four integrated acoustic receiver units and internal readout electronics. This design concept is pictured in \cref{fig:prototype-a}. It uses a P-ONE pressure housing hemisphere that consists of a \unit[17]{"} glass hemisphere attached to a titanium flange using deepsea-applicable epoxy resin, and a custom Delrin backing with a penetrator for cabled operation.To fix the orientation of the buoyant glass hemisphere, four rope segments are fixed to the Delrin back piece and connected to a single \unit[40]{lbs} weight. When submerged along with the module, this keeps the glass hemisphere stably pointing downwards. The four acoustic receivers were installed with planar offset angles of about \unit[90]{degrees}. An assembled module is pictured in \cref{fig:prototype-c}.
\par
\begin{figure}[h!]
    \centering
    \includegraphics[width=.94\textwidth]{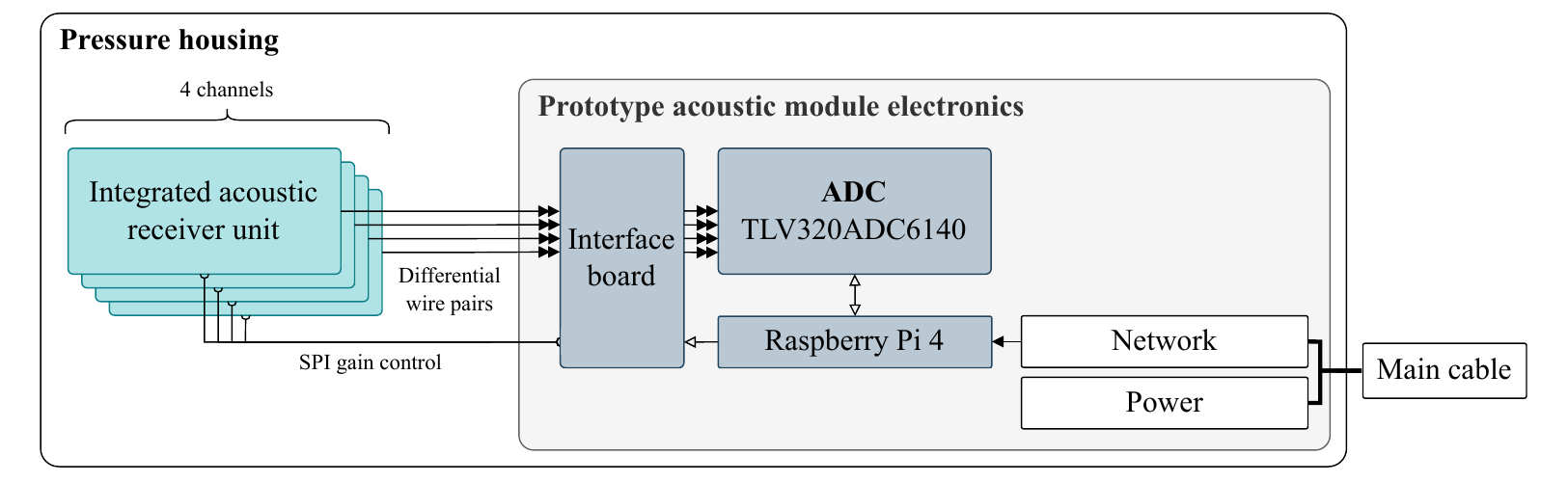}
    \caption{Logic diagram of the data acquisition concept for the P-ONE prototype module readout.}
    \label{fig:prototype-logic}
\end{figure}

\noindent
For digitization, our prototype module uses the integrated ADC6140EVM-PDK evaluation board with a custom I$^2$C interface to a Raspberry Pi 4. This allows setting up, controlling, and taking data with the ADC. The readout software is based on a public software package~\cite{mulier_audio-recording-firmware-raspi-tlv320adc6140_2022}, which provides ADC device overlays for the Raspberry Pi kernel and Python-based interface and readout code. A custom electronics board connects the four receivers to power and SPI communication with the Raspberry Pi, and allows SPI-control of the receiver gain. Network connection and power is provided via the main cable. The logic diagram of the prototype module electronics and readout is pictured in \cref{fig:prototype-logic} and is a near-identical clone of the anticipated P-ONE readout system.
%

\section{Design characterization}
\label{subsec:optimization}
The design of the mechanical and electronic design of the receiver was optimized in the laboratory. This included the selection of the piezo-acoustic element and the coupling epoxy, the design of the mechanical housing, and the optimization of the filtering and amplification electronics. In a second step, the integrated sensitivity of the receiver was tested within a sea water test tank located at the Marine Technology Center (MTC), operated by ONC. This key quantity of the prototype module determines what level of external acoustic pressure amplitudes it is able to pick up over its intrinsic noise. Furthermore, it determines the maximum detection distance of a source with a given output sound pressure amplitude and emission pattern.

\subsection{Experimental setups}

\begin{figure}[h!]
    \centering
    \begin{subfigure}{.45\textwidth}
      \centering
      \includegraphics[width=.95\textwidth]{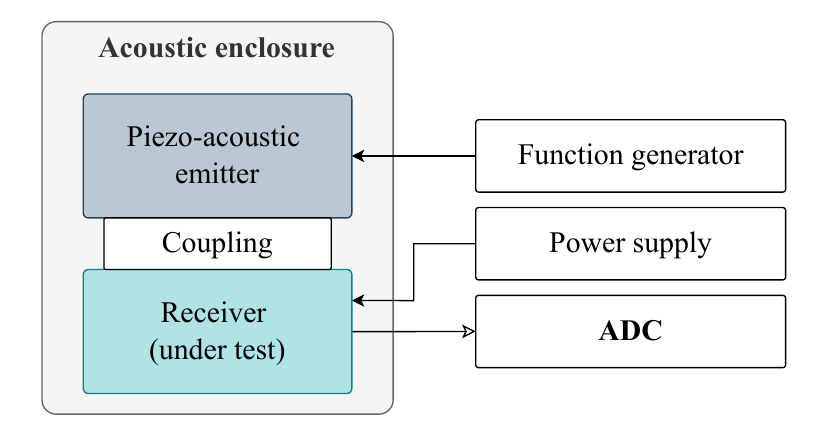}
      \caption{Laboratory measurement concept.}
      \label{fig:setup-lab-a}
    \end{subfigure}%
    \hspace{1mm}
    \begin{subfigure}{.53\textwidth}
      \centering
      \includegraphics[width=.95\textwidth]{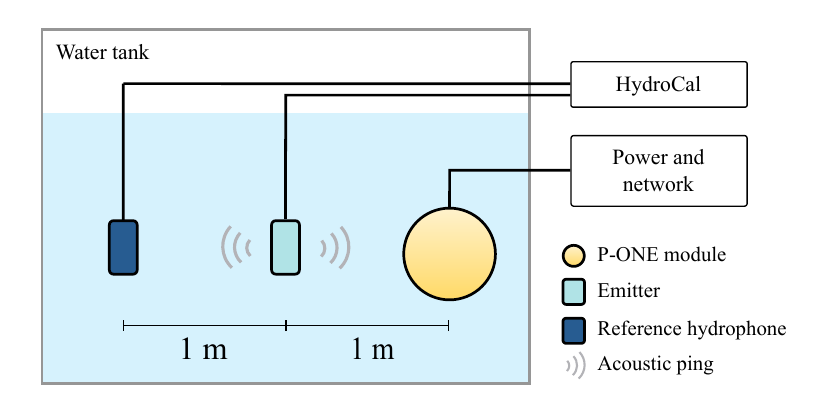}
      \caption{Water tank measurement concept.}
      \label{fig:setup-lab-b}
    \end{subfigure}
    \caption{\textbf{a)} Laboratory measurement setup showing a function generator, power supply, and an oscilloscope for operating piezo-acoustic emitters and receivers. Mechanical coupling was either direct or through glass. \textbf{b)} Water tank setup for absolute sensitivity measurements showing the prototype module and the HydroCal system in the chilled sea water test tank.}
    \label{fig:setup-lab}
\end{figure}

\paragraph{Laboratory} We used a laboratory measurement setup o optimize the design and performance of the acoustic receiver in air. The setup, pictured in \cref{fig:setup-lab-a}, consisted of an acoustically-dampened enclosure that contained a piezo-acoustic emitter coupled to a receiver. Signals with variable frequency and amplitude were produced with a function generator and read out with an oscilloscope (Rohde \& Schwarz RTM3004) and a power supply (Keithley 2230-01). The frequency test range was \unit[1]{Hz} to \unit[100]{kHz}.

\paragraph{Water tank} The prototype module's sensitivity was measured using the HydroCal system~\cite{biffard_integrated_2022} within the sea water test tank located at the Marine Technology Center (MTC) operated by ONC. This system uses a calibrated reference hydrophone to measure the sensitivity of a device by observing acoustic signal chirps of a common emitter with various, monotonic frequencies in both. This setup is shown in \cref{fig:setup-lab-b} and uses an equidistant acoustic emitter to the calibrated sensor and the one under test. Frequency ranges between approximately \unit[$1 - 100$]{kHz} can be calibrated this way. The absolute sensitivity of the sensor of interest (in units of V$^2/$$\mu$Pa$^2$) is obtained by comparing its measured response with that obtained by the calibrated reference hydrophone scaled to absolute sound pressure level through its calibration.
\par\noindent
Orientation dependence of the different channels in the prototype module is expected as the different acoustic receiver units are installed in \unit[90]{degree} increments to provide a uniform coverage of \unit[360]{degrees}. This effect can not be compensated in-situ, so the measurement is carried out over a fixed angular orientation of the sphere relative to the acoustic emitter and measurements for all channels are taken. 

The HydroCal system was controlled using a dedicated DAQ system and computer which was responsible for controlling the acoustic pulse emission and data taking of the calibrated reference hydrophone. Synchronously to a HydroCal measurement being run, data with the P-ONE prototype module is recorded. Using integrated synchronization pulses within the emitted pulses of the HydroCal run, both data can be joined in an offline processing step later-on. 
%
%

\subsection{Results}
Several iterations of the acoustic receiver design were developed and optimized. This targeted the mechanical design, coupling, and electronics performance. 

\paragraph{Mechanical coupling}
The mechanical coupling of the piezo-acoustic element to the glass interface is critical for acoustic signal pick-up. Several techniques for behind-glass piezo-acoustic element coupling were investigated for the AMADEUS~\cite{antares_collaboration_amadeus_2011} and KM3NeT ~\cite{viola_acoustic_2016} experiments, and the IceCube Upgrade~\cite{turcotte_development_2019,heinen_acoustic_2021}. 
These studies agree that epoxy is a suitable integration technique, and while mechanical coupling can be improved with more intricate designs, its ease of assembly makes it well suited for large-scale installations. Different epoxy resins were tested for application in P-ONE, with hard-curing epoxies resulting in a stronger signal pickup due to their acoustic impedance similarity to glass. As a result of these tests, we decided to use the two-part rigid-curing epoxy "The Original Cold Weld" made by J-B Weld~\cite{noauthor_j-b_nodate}.

\begin{figure}[h!]
    \centering
    \includegraphics[width=.95\textwidth]{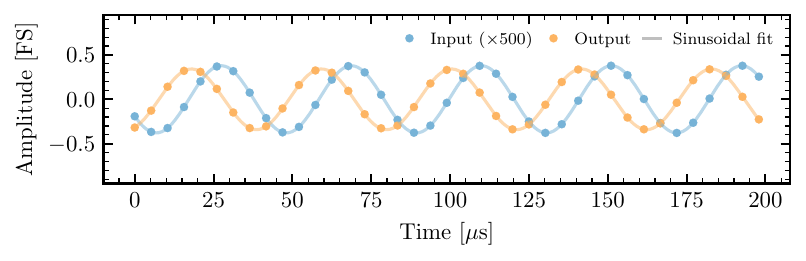}
    \caption{Example signal response of the electronics stack given a \unit[24.2]{kHz} input sine wave with \unit[2]{mV} amplitude in full-scale (FS) units of the digitizer. Data points represent digitizer samples.}
    \label{fig:example-waveform}
\end{figure}

\paragraph{Filtering and amplification}
The filter- and programmable pre-amplifier boards together form the electronics stack that filters and amplifies the raw signals of the piezo-acoustic element. The use of active filters and pre-amplifiers results in frequency-dependent bandwidth and phase shifts. Noise introduction through the unshielded transmission of signals and through the voltage supplies is a further concern. An example waveform of a raw \unit[24.2]{kHz} sine wave passed through the electronics stack and its phase-shifted, amplified response is given in \cref{fig:example-waveform}.
\par\noindent

\begin{figure}[h!]
    \centering
    \includegraphics[width=.95\textwidth]{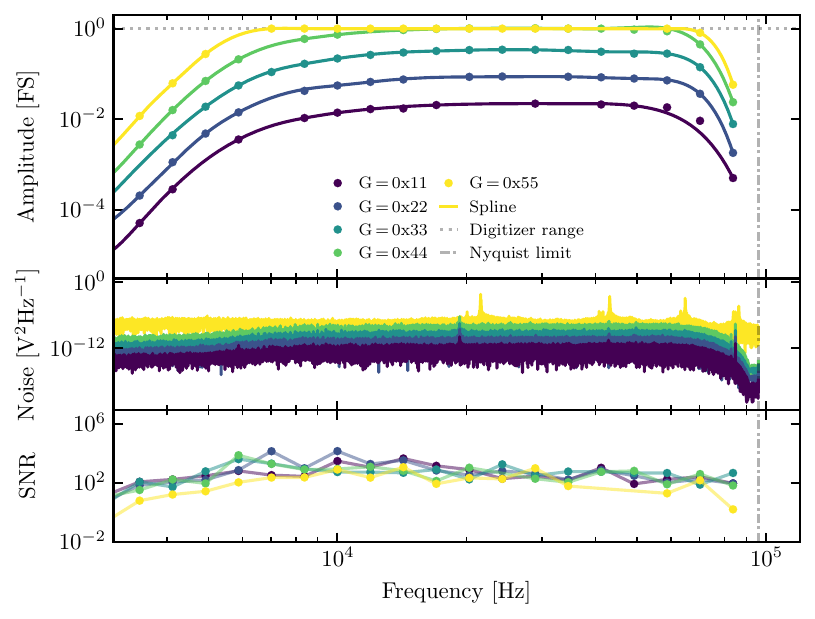}
    \caption{Bandwidth response in full-scale digitizer units (top), noise power spectral density (center) and the extracted signal-to-noise ratio (SNR) of the electronics stack as a function of frequency (bottom) at various programmed gain values (color). Data points represent sinusoidal fit amplitudes and lines are cubic splines. All measurements were done using sinusoidal input pulses of \unit[2]{mV} amplitude and the programmed, hexadecimal PGA gains range from $g \simeq 1 - 256$~\cite{linear_technology_corporation_ltc6912_2004}.}
    \label{fig:bode}
\end{figure}

In combination, the filter and pre-amplifier stack combine multiple stages of limited bandwidth. After passing the stack, the end-to-end signal-to-noise ratio (SNR) and bandwidth of the system can be measured using sine wave input signals of varying frequency and the ADC. Using input and output waveforms as shown in \cref{fig:example-waveform}, the electronic performance can be quantified. The results of these measurements for a prototype receiver stack are shown in \cref{fig:bode} as a function of frequency and shows the expected stable bandwidth for frequencies between \unit[$10-50$]{kHz}, the filter roll-off below \unit[10]{kHz}, the noise power spectral density (PSD), and the SNR of the system as a function of gain. Here, the SNR is the ratio of PSD amplitudes in signal and noise measurements.
\par\noindent

\begin{figure}[h!]
    \centering
    \includegraphics[width=.95\textwidth]{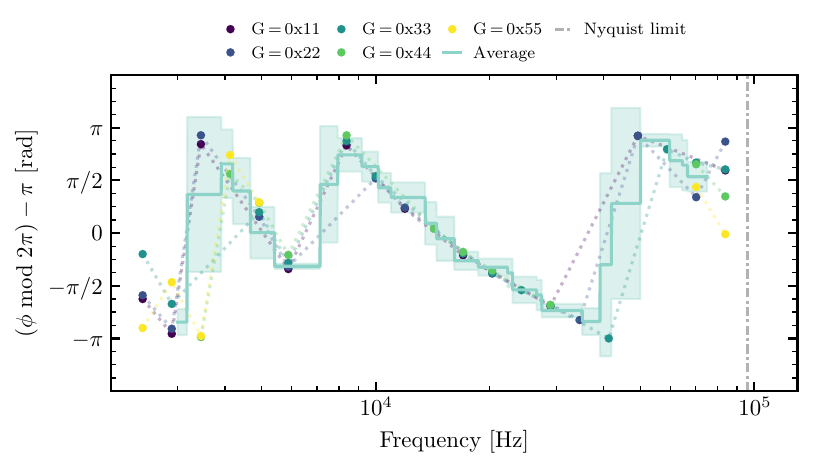}
    \caption{Phase shift of the full electronics stack averaged over different values of the programmable gain. The image shows the measured phase (modulo \unit[2]{$\pi$}) of the continuous sinusoidal wave as a function of frequency. A nearest-neighbor interpolation that averages the phase response of the different gain configurations is also shown.} 
    \label{fig:phase}
\end{figure}

In our prototype, the maximum usable gain of the programmable stage was observed to be $G\sim 512$ where the observed SNR in the ADC approaches~$\sim 1$. This results in a maximum usable system gain from piezo-acoustic element to signal output of \unit[$G\sim 25 \times 512 \times 2 = 25\,600$]{V/V}, which includes the differential output stage with a gain of $G\sim 2$ and the initial filter stage. The phase shift introduced is shown in \cref{fig:phase}. It displays the expected characteristics for high-order active filters~\cite[e.g.][]{hoang_optimized_2009,zumbahlen_phase_nodate}. Knowledge about this shift is critical for precision positioning as it will skew the physical arrival time of acoustic pings and needs to be corrected for. 

\paragraph{Integrated sensitivity} We conducted several HydroCal runs with the P-ONE prototype module over a frequency range of \unit[$1-50$]{kHz} and for various gain settings of the programmable pre-amplifier within the piezo-acoustic receiver stack. The goal of this campaign was the measurement of the absolute sensitivity of all four channels in the P-ONE prototype module as a function of frequency and gain. The data of both systems is analyzed offline with software included within the Hydrocal framework~\cite{biffard_integrated_2022}. Measured data files for the reference hydrophone, the sensor of interest, and the reference hydrophone calibration are input to the software. The algorithm searches for the synchronization pulses in both files, aligns them, and then measures the responses of all monotonic frequency signals in both sensors. The sensitivity as a function of frequency is then calculated by iteratively comparing magnitudes of fast Fourier transform spectra between reference and prototype module data. After scaling the prototype module results with the absolute sensitivity from the reference calibration, the output is the absolute sensitivity in units of V$^2/$$\mu$Pa$^2$. 

\begin{figure}[h!]
    \centering
    \includegraphics[width=0.95\textwidth]{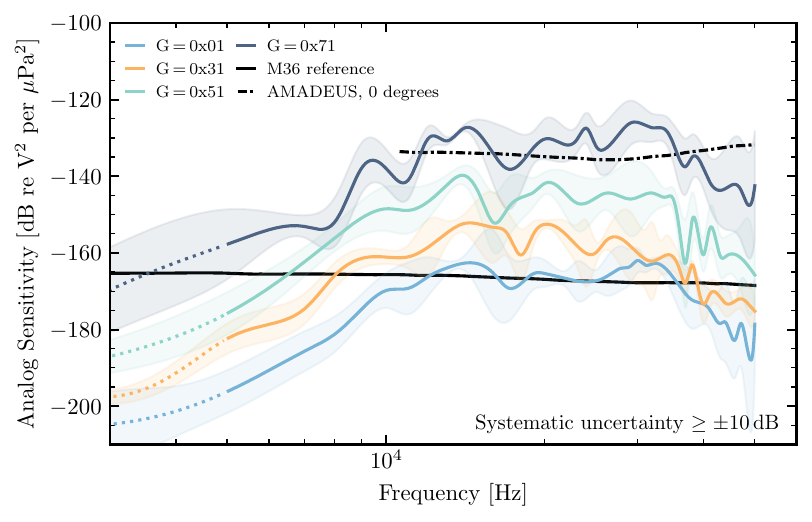}
    \caption{Absolute sensitivity measurement of a P-ONE prototype module obtained with the HydroCal system. The image shows the absolute sensitivity averaged over all piezo-acoustic receiver channels of the P-ONE prototype module as a function of frequency and for different gain settings of the programmable pre-amplifier. The frequency range from \unit[$1-5$]{kHz} was dominated by noise in the acoustic receiver units and reported sensitivities of the HydroCal software are not reliable. The calibration data of the reference hydrophone M36 and a sensitivity measurement of a similar acoustic module for the AMADEUS experiment~\cite{antares_collaboration_amadeus_2011} are also shown. The systematic uncertainty is contributed to the background noise noted during background runs.}
    \label{fig:sensitivity}
\end{figure}

The results of this measurement campaign for the P-ONE prototype module are summarized in \cref{fig:sensitivity} and show the absolute sensitivity averaged over all piezo-acoustic receiver channels. Between \unit[$10-40$]{kHz} a relatively uniform sensitivity of approximately \unit[-165]{dB re V$^2/$$\mu$Pa$^2$} and \unit[-125]{dB re V$^2/$$\mu$Pa$^2$} is observed for the lowest and highest possible gain setting, respectively. For these configurations, our result range is comparable with the M36 reference hydrophone of the HydroCal system~\cite{biffard_integrated_2022} and a similar piezo-type instrument built for the AMADEUS experiment~\cite{antares_collaboration_amadeus_2011}. A drop in sensitivity can be observed in the range of \unit[$5-20$]{kHz} and \unit[$40-50$]{kHz}, reaching minimum values of approximately \unit[$-200$]{dB re V$^2/$$\mu$Pa$^2$} and \unit[$-180$]{dB re V$^2/$$\mu$Pa$^2$} in the two ranges, respectively. Noise dominated the frequency range below \unit[5]{kHz} and reliable sensitivity estimates were not possible. With the integrated filter rolling off steeply below \unit[10]{kHz}, this behavior is expected. The limited bandwidth of our electronics also explain the sensitivity roll-off at high frequencies.
\par\noindent
%
\section{Positioning field test}
\label{subsec:positioning}
The final test of the prototype module was its positioning performance, which is needed for array geometry monitoring in P-ONE. A field test was carried out on the Saanich Inlet near Sidney, BC, using a small ONC-operated ship as shown in \cref{fig:setup-boat}. \textcolor{rev1}{The goals of the test are to exercising the entire system in an environment similar to P-ONE, and to have a first look at in-situ data and noise.}

\subsection{Experimental setup}
Three autonomous acoustic pingers (APs) by Sonardyne (WMT 8190-3111), each with a unique frequency peak within \unit[$24 - 30$]{kHz}, were deployed and positioned into a triangular pattern on the seafloor of the inlet. This triangle is approximately equilateral with a distance between center and vertex of about \unit[200]{m}, and each position was recorded using the ship's GPS system. The pingers, each rigged with ropes to a weight and a float, were then lowered to the seafloor at a depth of approximately \unit[165]{m}. The float kept the pingers approximately \unit[10]{m} off the ground. The P-ONE prototype module was cabled to the ship and connected by rope to an acoustic interrogator unit (Sonardyne {WMT 8190-3111, upgraded to an interrogation unit) and a weight. The distance between the interrogator and the prototype module was approximately \unit[10]{m}, and both were cabled to dedicated computers on the ship for control and data acquisition.
\begin{figure}[h!]
    \centering
    \includegraphics[width=0.8\textwidth]{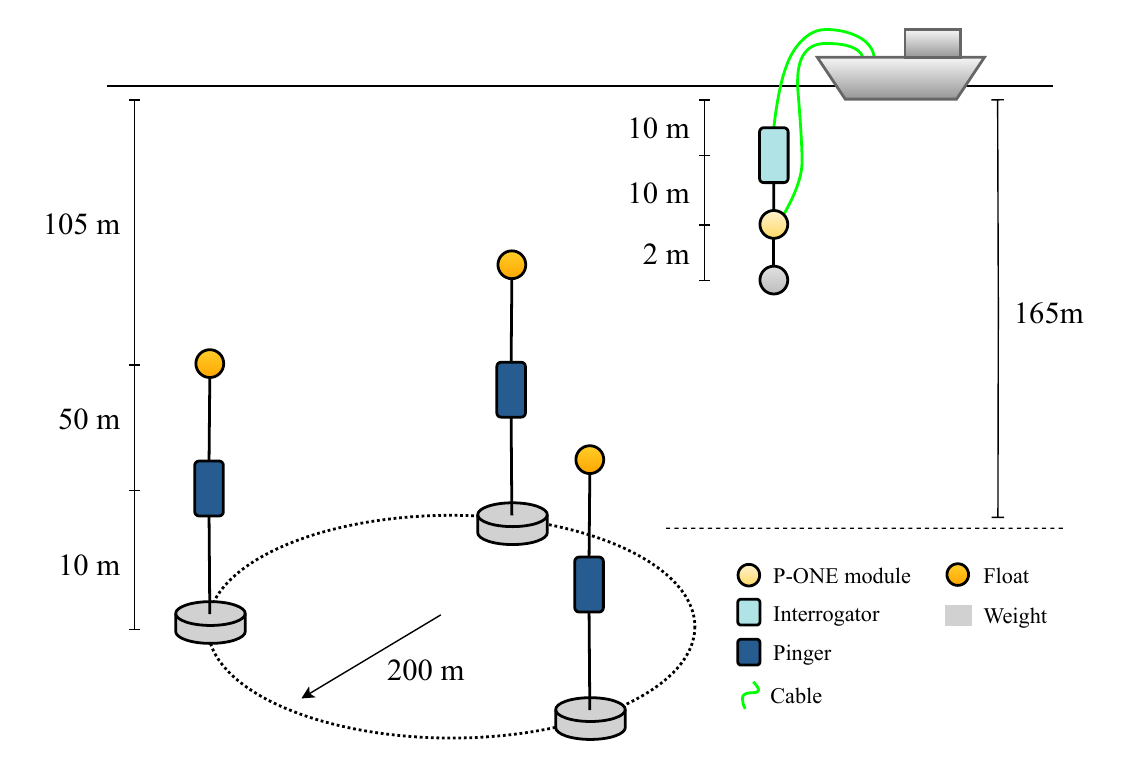}
    \caption{Field test measurement setup for the P-ONE prototype module. Three autonomous pingers were installed \unit[10]{m} above the bottom of Saanich Inlet and kept upright with floats. The interrogator unit and the P-ONE module were rigged together with a distance of \unit[8]{m} and lowered into the water from a ship at various surface positions. Interrogation pulses of the interrogator triggered acoustic responses from all pingers while the P-ONE module was recording acoustic data.}
    \label{fig:setup-boat}
\end{figure}\par\noindent
At each position pictured in \cref{fig:positions}, the cabled interrogator-prototype rig was lowered about \unit[20]{m} into the water, and a serial connection to the interrogator allowed setting its configuration as well as triggering acoustic signals of all pingers. Special serial commands sent to the interrogator unit allowed remote acoustic programming of the configuration in all pinger units through the acoustic Sonardyne communication protocol. The P-ONE prototype module cable carried power and network connection, and the module was controlled by a separate computer.
\begin{figure}[h!]
    \centering
    \includegraphics[width=0.9\textwidth]{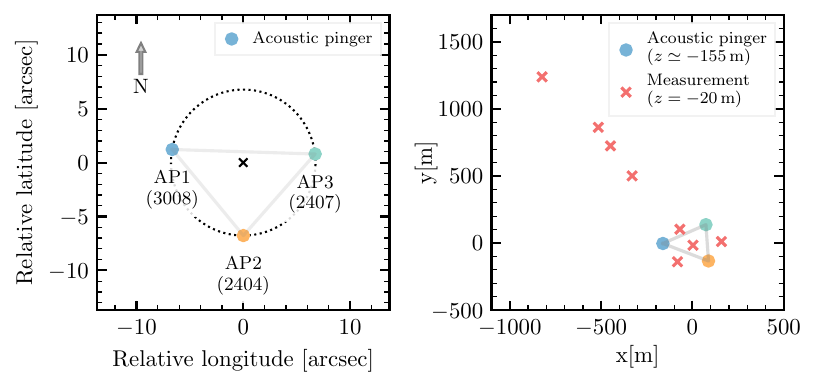}
    \caption{Coordinate maps for the measurement campaign of the P-ONE prototype positioning system. The left figure shows a detailed view of the latitude and longitude coordinates of the ship location where the acoustic pingers (APs; 4-digit unique ID) were deployed, centered around their circumcircle (dashed). The right figure shows a birds-eye view of the Cartesian coordinate plane after transformation with the WGS84 dataset~\cite{gmbh_httpswwwklokantechcom_wgs_nodate}, and includes the various measurement locations where data was collected. The pingers were installed in an approximate depth of \unit[165]{m} and the measurements were taken with the P-ONE prototype module at a depth of \unit[20]{m}. The uncertainty on the pinger location relative to the ship's GPS is estimated to be at least \unit[10]{m}.}
    \label{fig:positions}
\end{figure}\par
In the beginning of each position measurement, first the emitting power and turn-around time of all pingers were programmed acoustically through the interrogator. Then, the P-ONE prototype module starting taking data throughout a series of periodic acoustic interrogations. The pingers responded to these interrogations with their own acoustic pings after the programmed turn-around time, which were programmed to \unit[$240, \, 440, \, 640$]{ms} for AP1, AP2, and AP3, respectively. This cycle was monitored on the ship trough the serial connection with the interrogator unit, and the prototype module was taking several minutes of data while pings were being emitted. With the interrogator close to the P-ONE module, the starting time of the acoustic cycle is recorded in addition to the acoustic pinger responses. The repetition frequency of the periodic time series sent out by the interrogator was set to three seconds. Considering that the signal has to travel the physical distance twice (from the interrogator to the pinger and back), this theoretically allows measuring acoustic pulses from distances of more than \unit[2]{km} without overlap between cycles~\footnote{Assuming a sound speed in ocean water of approximately $v_s\sim\;$\unit[1500]{m/s}.}. 


\subsection{Results}
\paragraph{Acoustic pulse identification} At each position, the P-ONE prototype module took several minutes of data in increments of \unit[30]{s} length during the periodic pinging cycle of the interrogator. The next critical step is the identification of acoustic pulses. In addition to the pinger signals, the test environment in the inlet was not noise free. Acoustic instrumentation and other ships caused transient acoustic noise that was recorded in addition to the interrogation cycles. Electronic noise in our system contributed noise in the kilohertz frequency ranges. Therefore, a multi-step algorithmic approach was needed to clearly identify acoustic pulses in data. First, we generate a Fourier-transform time series spectrogram of the data and set up a rectangular window with a time width equal to the acoustic pulse length of \unit[8]{ms} and a frequency height of \unit[20]{kHz}. We shift this window through the spectrogram and sum its power spectral density and run a peak-finding algorithm over this window sum time series to find the contributing peaks over background. Prior knowledge about the geometry allows identification of the individual pingers. A Fourier transform of each pulse window delivers the median frequency used to estimate its arrival time. The phase shift of the receiver unit is small and omitted here. An example of this is shown in \cref{fig:pulse-id}. 
\begin{figure}[h!]
    \centering
    \includegraphics[width=.9\textwidth]{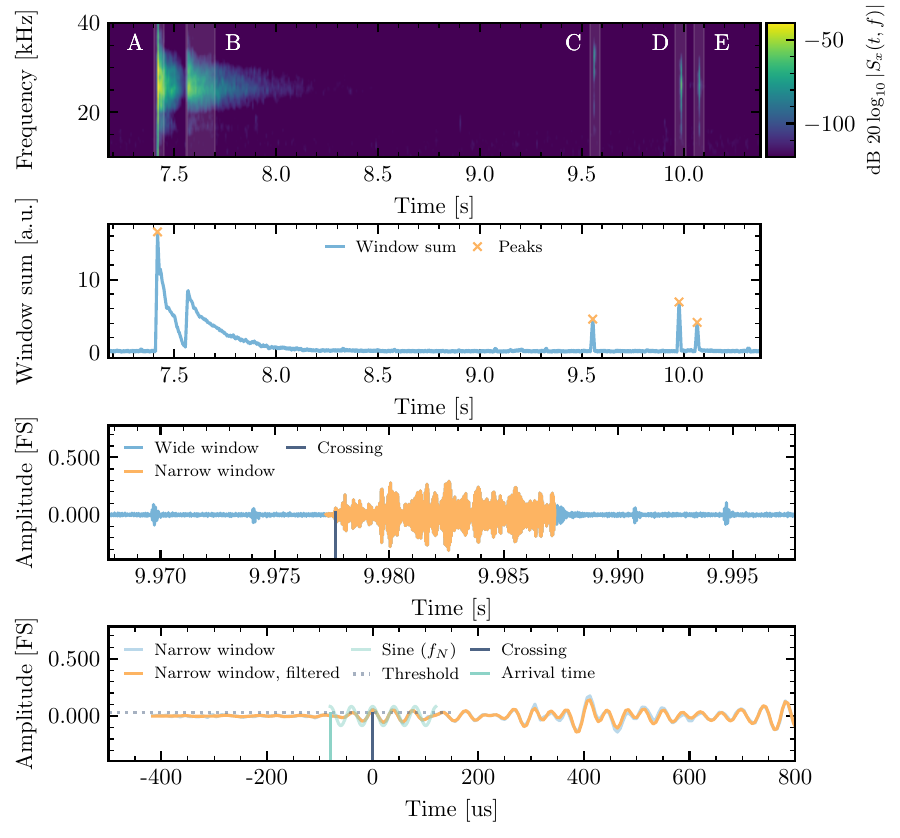}
    \caption{Acoustic pulse identification example at the furthest measurement location, approximately \unit[1.6]{km} from the pinger triangle. The first image shows the frequency spectrogram of one cycle in the periodic pinging series as a function of time. The labels indicate the sources of different arriving signal fronts: the interrogation pulse (A), its sea floor reflection (B), and the three distinct pinger responses (C,D,E). The second image shows the sum of a sliding window moving across the spectrogram and includes identified pings that pass the peak quality criteria. The third and fourth images show raw acoustic receiver data for pulse "C" in broad and detailed view as well as the arrival time extraction. For the latter, a dynamic threshold based on pre-pulse noise is used together with the main pulse frequency, $f_N$, to estimate the physical arrival time. The phase shift of the stack electronics is small compared to the peak-finding accuracy and is therefore omitted here.}
    \label{fig:pulse-id}
\end{figure}\par\noindent
%

\paragraph{Time of flight} 
Acoustic rays propagating through sea water follow Snell's law and bend toward smaller sound speeds. The propagation path and time of an acoustic wave is therefore directly correlated with the sound speed variation. The longer bending of paths taken by sound waves in a water column therefore depends on the geometric position of the emitter and receiver and the intermediate sound speed profile (SSP). In the Saanich Inlet, the SSP of the water column is measured by a vertical profiling system operated by ONC with publicly available data~\cite{ocean_networks_canada_society_yarrow_2024}. The measured values for the day of the test are shown in \cref{fig:ssp} as a function of depth. The median and \unit[$(10,90)$]{\%} quantile range is measured to be \unit[$\langle v_s \rangle = 1485.83^{+3.76}_{-1.01}$]{m/s} and \unit[$\langle v_s' \rangle = 1492.17^{+1.38}_{-1.13}$]{m/s} over depth ranges of \unit[$10-160$]{m} and \unit[$10-20$]{m}, respectively.
\begin{figure}[h!]
    \centering
    \includegraphics[width=0.9\textwidth]{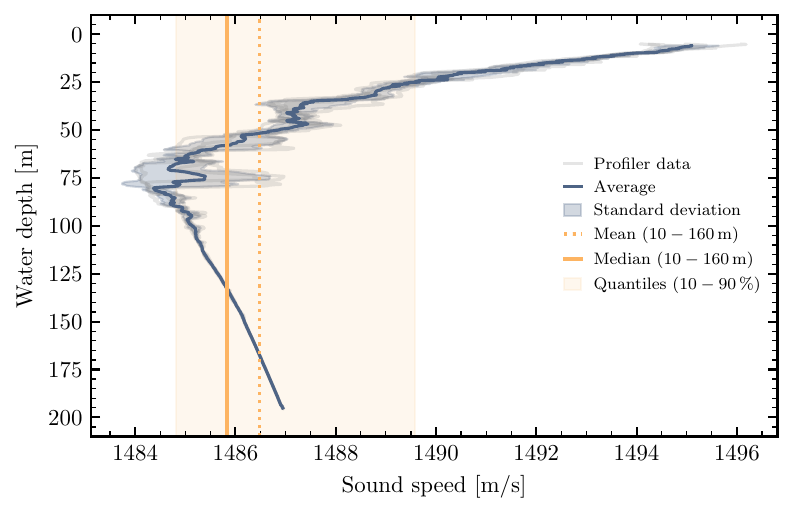}
    \caption{Sound speed profile of the Saanich Inlet as a function of water depth. The shown data are publicly available through ONC~\cite{ocean_networks_canada_society_yarrow_2024} and include four vertical profiles taken in six-hour intervals on the day of the test. The mean, median and \unit[10-90]{\%} quantiles for the sound speed at depths between \unit[$10-160$]{m} are also shown.}
    \label{fig:ssp}
\end{figure}\par\noindent

The simulation of acoustic rays propagating through water is done using the BELLHOP acoustic ray tracing framework~\cite{jensen_computational_1995,chitre_org-arlarlpy_nodate,noauthor_acoustics_nodate}. Using the expected positions of the emitters and receivers, we estimated the propagation time of bending acoustic rays as a function of distance. The results of an example ray-tracing scenario are shown in \cref{fig:ray-tracing} together with an estimate of the propagation time using the geometric distance and the median speed of sound. Generally, acoustic waves increasingly bend for longer distances and, in turn, spend more time in regions of increased speed of sound. This results in increasingly shorter travel times in reality than when estimating with the geometric distance and the median sound speed. The total correction amounts to \unit[$0.02 - 0.1$]{\%} deviation in the geometric distance range of \unit[$100-2000$]{m}. Furthermore, several ray reflection scenarios are present, however, reduced amplitudes and time delays make them a second-order effect.
\begin{figure}[h!]
    \centering
    \includegraphics[width=0.9\textwidth]{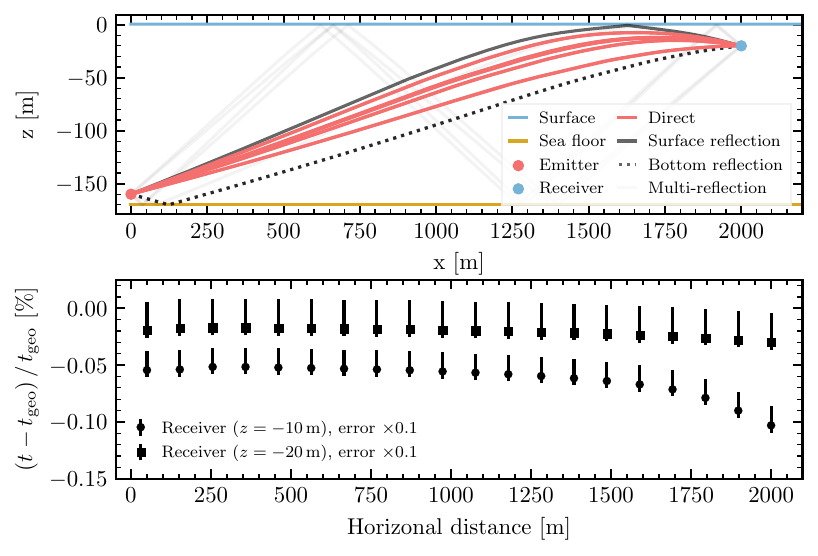}
    \caption{Acoustic ray-tracing using the BELLHOP framework. The top figure shows an example emitter/receiver realization and depicts the direct and indirect acoustic rays when propagated through the sound speed profile of the Saanich Inlet. Multiple ray solutions can be found to hit the receiver module, including direct rays bending in the sound speed profile and indirect rays reflected off the sea floor, the ocean surface, or both. The bottom figure shows the time of flight error for two receiver depths introduced by using the direct geometric path and the median sound speed of the relevant depth range ($t_\mathrm{geo}$) instead of a propagated ray ($t$). The errors represent the deviations resulting from the \unit[$(10,90)$]{\%} quantile ranges of the median sound speed assumption. A maximum deviation of \unit[0.1]{\%} is reached for a distance of \unit[2]{km} and a shallow-depth receiver.}
    \label{fig:ray-tracing}
\end{figure}\par\noindent

Both the prototype module and the Sonardyne system independently monitor the time of flights (TOFs) of the interrogation cycles. The geometric setup of both systems is shown in \cref{fig:propagation-geo} and, given the physical connection of both systems, maximal correlation between the two measurements is expected. For any given pinger, the TOF in the Sonardyne system includes twice the propagation time of the acoustic pulse ($t_\text{IP}$), from interrogator to pinger and back, as well as the turn-around delay time of the acoustic pinger ($t_\text{tat}$). By defining the start time of the initial interrogation signal ($t_0$) and the stop time of the acoustic pinger response ($t_1$), the TOF is
\begin{align*}
    t_\text{tof} &=  t_1 - t_0, \\
    &= 2 \, t_\text{IP} + t_\text{tat}.
    \numberthis\label{eq:tof-1}
\end{align*}
Here, the arbitrary start time $t_0$ does not influence the result since the measured TOF is the time difference between the emitted and received pulses and only depends on the propagation time and the programmed turn-around delay time. However, since the prototype module is only an external observer of the Sonardyne interrogation cycle and not time synchronized with the system, $t_0$ and $t_1$ have to be extracted from its data given the geometric setup of the system. 
\begin{figure}[t!]
    \centering
    \includegraphics[width=\textwidth,trim={0 12mm 0 10mm},clip]{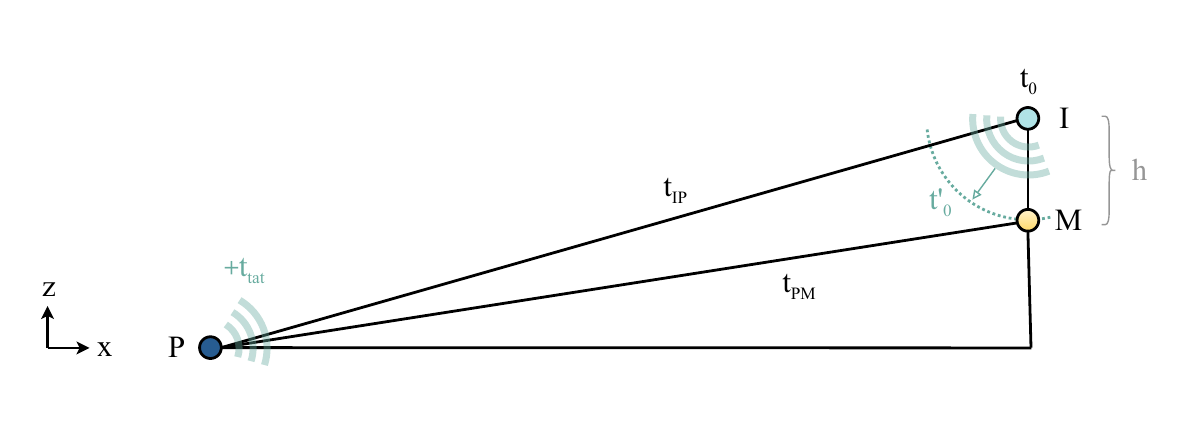}
    \caption{Two-dimensional sketch for the time of flight calculation for a given measurement location. The image shows the conceptual positions of the interrogator (I), the prototype module (M) and one acoustic pinger (P) as well as the ideal geometric paths of the acoustic signal propagation.}
    \label{fig:propagation-geo}
\end{figure}
For the prototype module, we see that for any given pinger, an effective TOF is observed that comprises
\begin{align*}
    t_\text{tof}' &= t_1' - t_0'\\
    &= t_\text{IP} + t_\text{PM} + t_\text{tat} - h / \langle v_s' \rangle \\
    &\simeq 2 \, t_\text{PM} + C + t_\text{tat} - h / \langle v_s' \rangle
    \numberthis \label{eq:tof-2}
\end{align*}
where the arrival time of the interrogation pulse at the prototype module $t_0'$ is related to $t_0$ through the known distance $h$ and speed of sound $\langle v_s' \rangle$ in the intermediate water, and $C$ is the distance-dependent difference between $t_\text{IP}$ and $t_\text{PM}$. The observables $t_0'$ and $t_1'$ are extracted from prototype module data. The set of distance-dependent TOF offsets $\{C_i\}$ between prototype module and Sonardyne for pinger $i$ can be computed using BELLHOP. This is shown in \cref{fig:tof-corr} as a function of lateral distance between emitter and receivers. Given the shallow slope of these results in the distance range of \unit[$1-1.75$]{km}, the GPS positions of the measurements can be used to estimate these offsets without introducing significant additional uncertainties. However, lateral sway between both receivers can change the offset and therefore adds systematic uncertainty.
\begin{figure}[h!]
    \centering
    \includegraphics[width=0.95\textwidth]{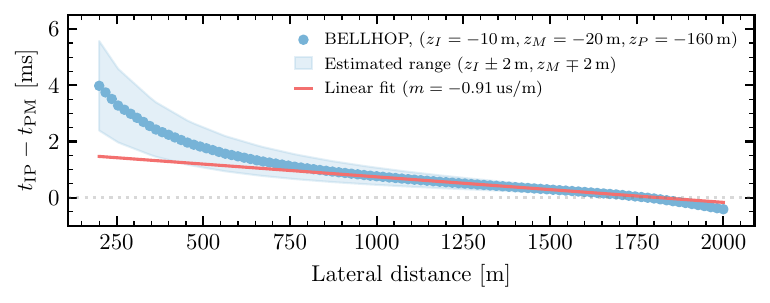}
    \caption{Difference in the ideal one-way ranging time from a pinger (P) to the interrogator (I) and the prototype module (M). Simulated with BELLHOP using the Saanich Inlet sound speed profile.}
    \label{fig:tof-corr}
    \vspace{-1\baselineskip}
\end{figure}\par\noindent
Altogether, this results in the correlation pictured in \cref{fig:sonardyne-corr}. There is excellent agreement between the two independent measurements to within the accuracy of our peak-finding algorithm of \unit[$230-280$]{$\mu$s} (or \unit[$35-42$]{cm} when assuming $v_s\simeq1500$m$/$s). It verifies that the prototype module, in combination with a simplistic peak-detection algorithm, is able to accurately track the ship's drift. Remarkably, it also highlights that this basic analysis can achieve a relative module position accuracy of \unit[$24-30$]{cm} in P-ONE when equipped with $N=2$ acoustic receivers and assuming that the accuracy of the reconstructed module center scales approximately as $\sigma/\sqrt{N}$. We further anticipate significant resolution improvements when combining more sophisticated peak-finding algorithms, receiver phase-shift calibration, and a mechanical model for all 20 modules spanning the P-ONE mooring lines~\cite{henningsen_pacific_2023}. Precise knowledge about pinger positions on the seafloor, however, remains essential.
\begin{figure}[h!]
    \centering
    \includegraphics[width=.95\textwidth]{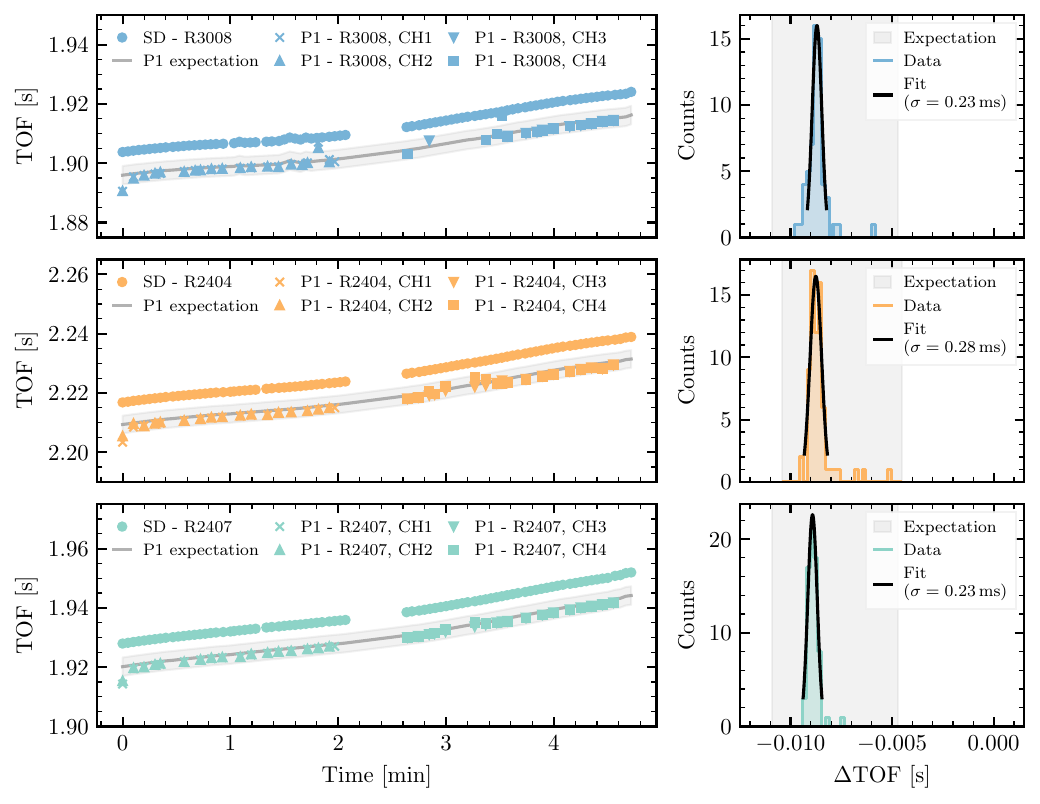}
    \caption{Time of flight measurements of both the Sonardyne and P-ONE system at the last measurement location with a distance of about \unit[1.5]{km} to the pingers. The left column of images shows the TOF measurement for each beacon as a function of time, corrected for its programmed turn-around delay time. It shows both the Sonardyne (SD) and P-ONE (P1) data with the latter broken into its receiver channels. Based on the geometry of the system, the Sonardyne data can be used to derive an expectation for the prototype module (continuous line with error band) where the uncertainty takes into account geometric uncertainties. The right column shows histograms of the difference between Sonardyne and P-ONE data. Normal distribution fits show standard deviations of \unit[$230-280$]{$\mu$s}, which lie within the accuracy of our pulse-finding algorithm. The expectation derived from geometry consideration and Sonardyne data is shown as a gray band.}
    \label{fig:sonardyne-corr}
\end{figure}

\paragraph{Positioning} Using the extracted TOFs we estimated the receiver positions. For this, a set of independent linear equations needs to be solved that describes the TOF measurements between the three pingers and the receiver. Using \cref{eq:tof-1,eq:tof-2}, we minimize 
\begin{equation}
    \chi^2 = \sum_j \left[ \sum_i \left( t_{ij} - \frac{w_{ij}}{\langle v_s \rangle} \, \sqrt{\left(x_j - x_i\right)^2 + \left(y_j - y_i\right)^2 + \left(z_j - z_i\right)^2} \right) \right]
    \label{eq:chi2}
\end{equation}
where $\vec{x_j} = (x,y,z)_j$ is the receiver coordinate vector at location $j$, $\vec{x_i} = (x, y, z)_i$ is the coordinate vector of pinger $i$, $\langle v_s \rangle$ is the median sound speed, $w_{ij}$ are the TOF corrections based on acoustic ray bending, and $t_{ij}$ are the measured one-way TOFs between receiver and pinger. 

We used the identified ranging times of both the prototype module and the Sonardyne system together with \cref{eq:chi2} to perform the positioning fit of the various measurement positions. Since both the receiver and pinger positions carry significant uncertainty, a Markov-chain Monte Carlo (MCMC) approach was chosen for the task. Prior uncertainties on all model parameters can be set up and the model posterior likelihood space is sampled to find a solution matching the measured TOFs in the presence of uncertainties. The PyMC package~\cite{abril-pla_pymc_2023} is used in combination with \cref{eq:chi2} to build and sample the MCMC model with priors listed in \cref{tab:mcmc}. We performed the fit using this model and the TOF data of the prototype module corrected for the one-way Sonardyne ranging time. To increase the robustness of the fit, we fit multiple positions and receiver channels simultaneously as the pinger positions are assumed to be static across different measurements. Given the large absolute uncertainties on the position of the pingers, the primary goals of this analysis are studying the fit's relative positioning performance and its correlation with GPS data.
\begin{table}[h!]
    \centering
    \begin{tabular*}{\textwidth}{@{\extracolsep{\fill}}lllll}
    \toprule
    \textbf{Model parameter} & \textbf{Prior} & \textbf{Prior parameters} & \textbf{Unit} \\
    \toprule
    Pinger position, $x$ & Normal & $\mu = x_\text{GPS}$, $\sigma=3$ & m\\
    Pinger position, $y$ & Normal & $\mu = y_\text{GPS}$, $\sigma=3$ & m \\
    Pinger position, $z$ & Normal & $\mu = z_\text{DSO}$, $\sigma=3$ & m  \\
    Receiver position, $x$ & Normal & $\mu = x_\text{GPS}$, $\sigma=10$ & m\\
    Receiver position, $y$ & Normal & $\mu = y_\text{GPS}$, $\sigma=10$ & m \\
    Receiver position, $z$ & Normal & 
    $\mu = -20$, $\sigma=3$ & m \\ 
    \bottomrule
    \end{tabular*}%
    \caption{MCMC model priors for the positioning fit using P-ONE TOF data. Priors on pinger and receiver positions use the surface location of the ship's GPS system and its depth sounder (DSO).}
    \label{tab:mcmc}
\end{table}\par\noindent
The relative coordinate system spanned by the three pinger positions determines the three measured TOFs to the receiver. These TOFs, however, are invariant under translations and rotations of the relative pinger coordinate system. Any uncertainty on the GPS-referenced pinger positions on the seafloor therefore directly translates into an absolute positioning uncertainty in the reconstruction. The absolute pinger locations on the seafloor are only lightly constrained by the surface GPS position taken immediately before their deployment. Furthermore, the ship's GPS antenna is about \unit[4]{m} away from its back-deck where the receiver was lowered into the water at each measurement location and its heading was unknown. That means, the receiver position is not equivalent but only correlated with the GPS position of the ship. Some relative shift between the reconstructed position and GPS in the lateral plane is therefore expected. These absolute positioning uncertainties are irreducible without a dedicated GPS-referenced calibration of the pinger positions.

Performing the fits, we find tantalizing agreement between the reconstructed position and the initial GPS position. This is shown in \cref{fig:reco-time} for the two measurement locations furthest from the pinger triangle, with average distances of about \unit[$1-1.5$]{km}. The ship drift in the water measured by GPS is consistent with the drift of our reconstructed position in time. The narrow relative spread of our reconstructed positions is furthermore consistent with the peak-finding accuracy apparent in \cref{fig:sonardyne-corr} and highlights the tracking performance of the prototype module. Single outliers in the fit are observed and are likely caused by failed identification of arrival time stamps in our peak-finding algorithm. These outliers are also evident in the TOF data shown in \cref{fig:sonardyne-corr} for the \unit[1.5]{km} location. 

In \cref{fig:reco-lateral}, multiple measurement locations are combined to quantify the fit's performance. For each location, a linear interpolation of the recorded GPS positions is compared to the reconstructed fit position as a function of measurement time. It is clear that the fit achieves good accuracy in between subsequent TOF measurements and the movement of the receiver relative to the ship can be traced throughout each measurement. This movement is not unexpected given the ship's drift in the wind during the measurement, and the single fix point of the receiver line at its back-deck. The aforementioned outliers are also clearly visible here. Overall, the fits can trace the GPS-tracked positioning with standard deviations of \unit[$0.8 - 2.2$]{m}. 
\begin{figure}[h!]
    \centering
    \includegraphics[width=0.95\linewidth]{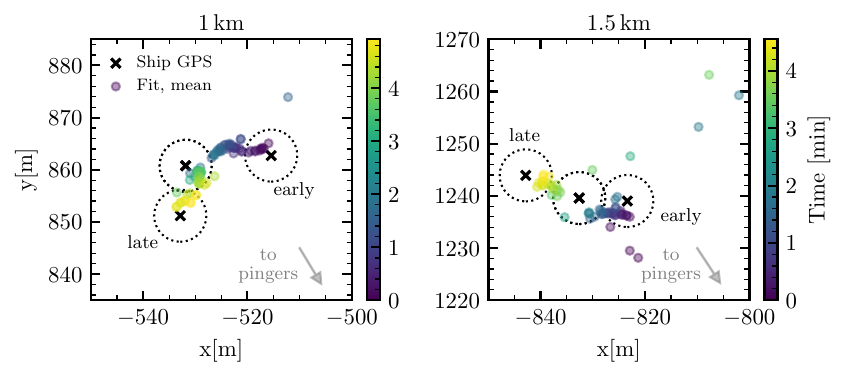}
    \caption{Time-resolved reconstruction of the prototype module position using our MCMC model at two measurement locations with an average distance to the pingers of about \unit[$1$ and $1.5$]{km}. Each point corresponds to the posterior mean of the fit given a single TOF measurement in a single receiver channel. The color of the point corresponds to the time of  recording. During data taking, two channels were recording simultaneously. The ship GPS location was recorded at the beginning, during, and at the end of the measurement cycle. The positional drift with time can be similarly observed in both the GPS and the reconstructed position fits using TOF data. Outliers in the fit are likely caused by erroneous arrival time stamping of our peak-finding algorithm and are also apparent in the raw TOF data shown in \cref{fig:sonardyne-corr}. For each point, the fit posterior distribution shows standard deviations of \unit[$3-4$]{m}, which are not shown for visual purposes.}
    \label{fig:reco-time}
\end{figure}\par
\begin{figure}[h!]
    \centering
    \vspace{-0.5\baselineskip}
    \includegraphics[width=0.95\linewidth]{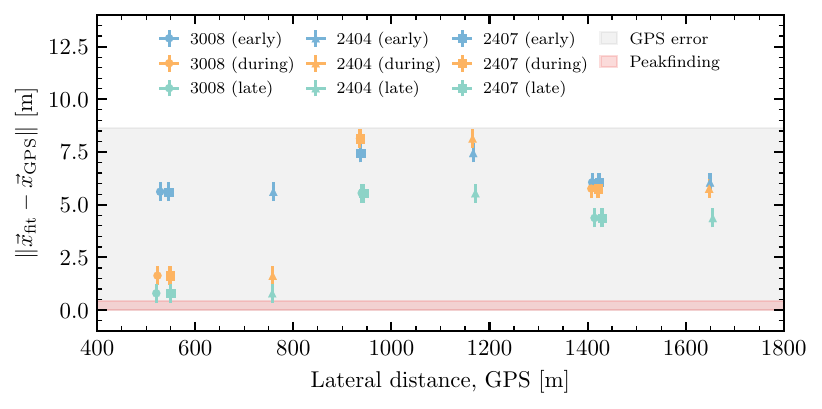}
    \caption{Difference between reconstructed and GPS position for three different measurement locations with an approximate pinger distance of \unit[$0.5-1.5$]{km}. The x-axis shows the GPS-based lateral distance between the pinger and the ship and the y-axis shows the euclidian distance between the fit position $(\vec{x}_{\text{fit}})$ and boat GPS position $(\vec{x}_{\text{GPS}})$. The gray error band shows the estimated GPS uncertainty and the error bars as well as the red error band depict the relative error based on the peak-finding accuracy. The offset from zero is likely caused by a systematic displacement between the GPS and the assumed positions for pingers and receiver.}
    \label{fig:reco-lateral}
\end{figure}\par\noindent

\section{Summary}
\label{sec:summary}
In this work, we present the design, characterization, and performance verification of the prototype positioning system for the Pacific Ocean Neutrino Experiment. This system combines commercial, autonomous acoustic pinger instruments, and a custom piezo-acoustic receiver design with filtering, amplification, and digitization electronics. A detailed characterization of the electronics design is performed in the laboratory, and the absolute sensitivity of integrated acoustic receivers in P-ONE instrument housings is characterized in a water test tank. We observe a uniform bandwidth response of our analog receiver electronics between \unit[$10-40$]{kHz}, and a relatively uniform absolute sensitivity in the same frequency range of approximately \unit[-165]{dB re V$^2/$$\mu$Pa$^2$} and \unit[-125]{dB re V$^2/$$\mu$Pa$^2$} for low and high programmable gain settings, respectively. 

In addition, we performed a positioning measurement in the sea water of the Saanich Inlet using three autonomous acoustic pingers and an acoustic interrogator by Sonardyne. This system closely resembles the acoustic positioning system that will be deployed in P-ONE. A peak-finding algorithm was developed to identify acoustic pings in piezo-acoustic receiver data and ultimately extract ranging times to the acoustic pingers. From this data, we demonstrate accurate tracking of the ship's movement which maximally correlates with the independent ranging measurement of the Sonardyne system to within the accuracy of our arrival time algorithm of \unit[$230-280$]{$\mu$s}. We further perform a successful three-dimensional reconstruction of the receiver position based on TOF that closely correlates with ship's drift visible in the TOF measurement. The relative reconstruction within the pinger coordinate system is accurate, however, the absolute position of our reconstruction within the GPS reference frame is dominated by the positioning uncertainties of the pingers. 

For P-ONE, the results of this basic analysis already indicate a relative positioning accuracy of \unit[$24-30$]{cm} for single measurements, which will be significantly boosted by more sophisticated peak-finding algorithms, phase-shift calibration, the introduction of a mechanical line model combining all 20 modules on the P-ONE mooring lines, and multiple measurements in quick succession because of the slow expected line movement. Separate  calibration campaigns are planned with dedicated, high-precision instrumentation~\cite{farrugia_northern_2019,hutchinson_initial_2022} to precisely locate acoustic pinger positions on the seafloor which ultimately translates our relative positioning accuracy into the absolute GPS reference frame.

In conclusion, we demonstrated a successful ocean measurement campaign using a small-scale realization of the P-ONE acoustic system. This furthermore encompasses a novel custom piezo-acoustic receiver design and included filter- and amplifying electronics which will be installed in the first line of P-ONE planned for deployment later this year. While more work on improving the TOF extraction is needed to increase precision, this first campaign shows promising performance and verifies the applicability of the proposed acoustic system design in P-ONE.

\section*{Acknowledgments}
\noindent We thank Ocean Networks Canada for the very successful operation of the NEPTUNE observatory, as well as the support staff from our institutions without whom this experiment and P-ONE could not be operated efficiently.  We acknowledge the support of the Natural Sciences and Engineering Research Council of Canada (NSERC) and the Canadian Foundation for Innovation (CFI). This research was enabled in part by support provided by the BC and Prairies DRI and the Digital Research Alliance of Canada (alliancecan.ca). This research was undertaken thanks in part to funding from the Canada First Research Excellence Fund through the Arthur B. McDonald Canadian Astroparticle Physics Research Institute. P-ONE is supported by the Collaborative Research Centre 1258 (SFB1258) funded by the Deutsche Forschungsgemeinschaft (DFG), Germany. We acknowledge support by the National Science Foundation. This work was supported by the Science and Technology Facilities Council, part of the UK Research and Innovation, and by the UCL Cosmoparticle Initiative. This work was supported by the Polish National Science Centre (NCN).

\printbibliography

\end{document}